\documentclass[11pt,letterpaper]{article}

\usepackage[margin=1in]{geometry}

\usepackage[utf8]{inputenc} %
\usepackage[T1]{fontenc}    %
\usepackage{hyperref}       %
\usepackage{url}            %
\usepackage{booktabs}       %
\usepackage{amsfonts}       %
\usepackage{nicefrac}       %
\usepackage{microtype}      %
\usepackage{xcolor}         %
\usepackage{dsfont}

\usepackage{amsmath}
\usepackage{amssymb}
\usepackage{mathtools}
\usepackage{amsthm}
\usepackage{thmtools}
\usepackage{thm-restate}  %
\usepackage{makecell}
\usepackage{enumitem}
\usepackage{bbm}
\usepackage{subcaption}
\usepackage[ruled,vlined]{algorithm2e}
\DontPrintSemicolon
\SetKwInOut{Input}{Input}
\SetKwInOut{Output}{Output}

\usepackage{natbib}
\usepackage{parskip}
\usepackage{authblk}

\usepackage[capitalize, noabbrev]{cleveref}

\usepackage[toc,page]{appendix}      %
\usepackage{titletoc}

\usepackage{minitoc}
\usepackage{makecell}

\definecolor{Darkblue}{rgb}{0,0,0.4}
\hypersetup{colorlinks=true,
            citebordercolor={.6 .6 .6},
            linkbordercolor={.6 .6 .6},
            citecolor=Darkblue,
            urlcolor=blue,
            linkcolor=black}

\usepackage{booktabs,amsmath,tikz} %
\usepackage{subcaption} %

\newtheorem{theorem}{Theorem}[section]
\newtheorem{lemma}[theorem]{Lemma}
\newtheorem{claim}[theorem]{Claim}

\newtheorem{proposition}[theorem]{Proposition}
\newtheorem{definition}[theorem]{Definition}
\newtheorem{remark}[theorem]{Remark}

\newtheorem{assumption}[theorem]{Assumption}

\usepackage{multirow}

\crefname{lemma}{lemma}{lemmas}
\Crefname{Lemma}{Lemma}{Lemmas}

\crefname{ineq}{inequality}{inequalities}
\Crefname{Ineq}{Inequality}{Inequalities}
\creflabelformat{ineq}{#2{\upshape(#1)}#3} 
\creflabelformat{Ineq}{#2{\upshape(#1)}#3} 

\crefname{definition}{definition}{definitions}
\Crefname{definition}{Definition}{Definitions}
\crefname{prop}{proposition}{propositions}
\Crefname{Prop}{Proposition}{Propositions}
\crefname{assumption}{assumption}{assumptions}
\Crefname{assumption}{Assumption}{Assumptions}

\crefrangeformat{equation}{(#3#1#4) --~(#5#2#6)}
\crefmultiformat{equation}{(#2#1#3)}{, (#2#1#3)}{, (#2#1#3)}{, (#2#1#3)}
\crefname{lemma}{Lemma}{Lemmas}
\crefname{section}{Section}{Sections}
\crefname{subsubsubsection}{Section}{Sections}
\crefname{remark}{Remark}{Remarks}
\crefname{figure}{Figure}{Figures}
\crefname{table}{Table}{Tables}
\crefname{theorem}{Theorem}{Theorems}
\Crefname{theorem}{Theorem}{Theorems}
\crefname{algo}{Algorithm}{Algorithms}

\newcommand{\BR}{\mathsf{BR}}

\newcommand{\br}{\mathbb{R}}

\newcommand{\xst}{{\x}^{\star}}

\newcommand{\ip}[2]{\left\langle #1, #2 \right\rangle}

\newcommand{\Secref}[1]{\hyperref[#1]{Section \ref*{#1}}}
\newcommand{\Appref}[1]{\hyperref[#1]{Appendix \ref*{#1}}}

\newcommand{\cX}{\mathcal{X}}
\newcommand{\cY}{\mathcal{Y}}

\newcommand{\cA}{\mathcal{A}}
\newcommand{\cB}{\mathcal{B}}

\newcommand{\polytope}{P}

\newcommand{\squishlist}{
 \begin{list}{$\bullet$}
  { \setlength{\itemsep}{0pt}
     \setlength{\parsep}{3pt}
     \setlength{\topsep}{3pt}
     \setlength{\partopsep}{0pt}
     \setlength{\leftmargin}{1em}
     \setlength{\labelwidth}{1em}
     \setlength{\labelsep}{0.5em} } }

\newcommand{\squishlisttwo}{
 \begin{list}{$\bullet$}
  { \setlength{\itemsep}{0pt}
    \setlength{\parsep}{0pt}
    \setlength{	opsep}{0pt}
    \setlength{\partopsep}{0pt}
    \setlength{\leftmargin}{0em}
    \setlength{\labelwidth}{1.5em}
    \setlength{\labelsep}{0.5em} } }

\newcommand{\squishend}{
  \end{list}  }

\newcommand{\x}{\boldsymbol{x}}
\newcommand{\z}{\boldsymbol{z}}
\newcommand{\util}{\boldsymbol{u}}
\newcommand{\y}{\boldsymbol{y}}
\newcommand{\h}{\boldsymbol{h}}

\renewcommand{\u}{\mathbf{u}}

\newcommand{\ball}{\mathbf{B}}

\newcommand{\eps}{\varepsilon}

\newcommand{\tx}{\tilde{\x}}

\newcommand{\transpose}{\mathsf{T}}

\DeclareMathOperator*{\E}{\mathbb{E}}

\DeclareMathOperator*{\argmin}{\arg\!\min}
\DeclareMathOperator*{\argmax}{\arg\!\max}

\newcommand{\stackval}{\text{StackVal}}
\newcommand{\optwithinsubspace}{\mathsf{OptimizeWithinPolytope}}

\newcommand{\searchpolytope}{\mathsf{SearchForPolytopes}}
\newcommand{\binarysearch}{\mathsf{BinarySearch}}

\newcommand{\moveonestep}{\mathsf{MoveOneStep}}

\newcommand{\normal}{\mathcal{N}}
\newcommand{\spanspace}{\mathrm{span}}
\newcommand{\nullspace}{\mathrm{null}}
\newcommand{\simiid}{\overset{\text{iid}}{\sim}}

\newcommand{\vecone}{\mathbf{1}}
\newcommand{\veczero}{\mathbf{0}}
\newcommand{\augconstraints}{G}
\newcommand{\constraints}{H}
\newcommand{\avgx}{\overline{\x}}

\usepackage[suppress]{color-edits}
\addauthor{ky}{green!60!black}
\addauthor{na}{purple!60!black}

\title{Learning Local Stackelberg Equilibria \\from Repeated Interactions with a Learning Agent}

\author[1]{Nivasini Ananthakrishnan}
\author[2]{Yuval Dagan} %
\author[1]{Kunhe Yang} %

\affil[1]{UC Berkeley\\
\texttt{\{nivasini,kunheyang\}@berkeley.edu}}
\affil[2]{Tel Aviv University\\
\texttt{ydagan@tauex.tau.ac.il}}

\date{}

\begin{document}

\maketitle

\begin{abstract}
  Motivated by the question of how a principal can maximize its utility in repeated interactions with a learning agent, we study repeated games between an principal and an agent employing a mean-based learning algorithm. Prior work by~\citet{brown2024learning} has shown that computing or even approximating the \emph{global} Stackelberg value in similar settings can require an exponential number of rounds in the size of the agent's action space, making it computationally intractable. In contrast, we shift focus to the computation of \emph{local} Stackelberg equilibria and introduce an algorithm that, within the smoothed analysis framework, constitutes a Polynomial Time Approximation Scheme (PTAS) for finding an $\varepsilon$-approximate local Stackelberg equilibrium. Notably, the algorithm's runtime is polynomial in the size of the agent's action space yet exponential in $(1/\varepsilon)$—a dependency we prove to be unavoidable. 
\end{abstract}

\newcommand{\broracle}{\mathsf{BROracle}}
\newcommand{\liplb}{C}
\newcommand{\changedetect}{\mathsf{DetectBRChange}}
\newcommand{\avgy}{\overline{\mathbf{y}}}

\section{Introduction}

In repeated games, agents often have incomplete information about the game and resort to using a learning algorithm to optimize their utility over time. We consider a repeated game between a strategic player, referred to as the \emph{principal}, and a player who follows a learning algorithm, referred to as the \emph{agent}. When the principal anticipates the learning algorithm used by the agent, how should she adjust her strategy to maximize her utility over time? The answer depends on both the information available to the principal and the specific learning dynamics of the agent.

In this paper, we study the setting where the principal has no knowledge of the agent's utility function, and the agent employs a \emph{mean-based} learning algorithm. A \emph{mean-based} learner is a learning algorithm that selects its strategy based on the empirical distribution of the principal's past actions. Specifically, at each round, the learner plays an approximate best response to the time-averaged history of the principal's play. This class includes fundamental algorithms such as Fictitious Play, Follow the Regularized/Perturbed Leader, and Multiplicative Weights Update. In particular, we focus on Fictitious Play, where the agent \emph{exactly} best responds to the empirical distribution of the principal's actions up to the current round.

Given that the agent is a mean-based learner, what should the principal aim to achieve? A natural goal is to devise a strategy that maximizes her cumulative utility over time, among all possible strategies. This objective has been considered in Mechanism Design \citep{braverman2018selling, cai2023selling, rubinstein2024strategizing}, Contract Design \citep{guruganesh2024contracting}, Information Design \citep{lin2024persuading}, and general games \citep{deng2019strategizing, mansour2022strategizing, brown2024learning}. However, even when the principal has full knowledge of the agent's utility function, computing the optimal long-term strategy is NP-hard in general games \citep{assos2024maximizing, assos2024computational}.

An alternative benchmark is the \emph{Stackelberg equilibrium} of the one-shot game. If the principal knows the agent's utility function, she can compute a Stackelberg strategy in polynomial time and play it repeatedly. Since a mean-based learner best responds to the empirical distribution of the principal's past actions, this ensures that the agent selects the Stackelberg best response in each round, thereby approximately yielding the one-shot Stackelberg value to the principal in each round. When the principal does not have access to the agent's utility function, she could approximate the Stackelberg equilibrium using a \emph{best-response oracle}—an oracle that, given a mixed strategy of the principal, returns the agent's best response \cite{letchford2009learning, peng2019learning, blum2014learning}. If the only information available about the agent comes from observing its decisions over time, the principal may attempt to influence the agent's learning process to induce best responses. This approach has been analyzed in various settings \cite{haghtalab2022learning, haghtalab2024calibrated}. However, for the fundamental class of mean-based learners, reducing the problem to learning via a best-response oracle introduces an exponential overhead. That is due to the fact that mean-based learners remember the whole history of play, hence they cannot be easily manipulated to produce best responses. In fact, \cite{brown2024learning} established that approximating the Stackelberg equilibrium in this setting requires exponential time.

\subsection{Our contribution}

The impossibility result of \cite{brown2024learning} raises the question of what can be learned from interactions with a mean-based learning agent when the optimizer lacks direct access to the agent's utility function. A key challenge arises from the fact that mean-based learners are slow to forget—they adjust their strategy based on the cumulative history of play rather than responding to individual queries. As a result, querying their best response to widely varying mixed strategies is infeasible, since shifting the empirical distribution of past plays requires many rounds. This makes traditional approaches based on best-response oracles impractical.

Given these constraints, the optimizer can only infer information from how the agent gradually adapts its strategy over time. Consequently, the optimizer's best option is to perform a local search by making small adjustments to its historical play—modifying the empirical distribution of past actions incrementally—and observing the agent's response. Through this process, the optimizer can progressively optimize its utility. A natural target for such an optimization process is an approximate \emph{local equilibrium}, where the optimizer cannot significantly improve its utility through small deviations in its historical play. In this work, we study the problem of efficiently finding an $\epsilon$-approximate local equilibrium, leveraging the structure of mean-based learning dynamics to design an algorithm for this setting.

We present a PTAS: an algorithm whose iteration complexity is exponential in $1/\epsilon$ but polynomial in the size of the game, with each iteration running in polynomial time. Furthermore, we prove that this exponential dependence on $1/\epsilon$ is unavoidable. To achieve our runtime guarantees, we impose a few natural assumptions to prevent adversarial instances and to ensures that the vectors in the learner's utility matrix are in general position.

\paragraph{Technical Contributions.}  
Optimizing against a mean-based learner introduces several challenges:

\squishlist
    \item \textbf{Local versus Global Queries:}  
    Prior work reduced the problem of finding an approximate Stackelberg equilibrium via interactions with a learner to a query-based approach: given an optimizer's mixed strategy \(x \in \Delta(\mathcal{A})\), one can query the learner's (exact or approximate) best response \(y \in \mathrm{BR}(x)\). However, for mean-based learners, queries are expensive if the mixed strategies differ substantially. To query \(\mathrm{BR}(x)\) and then \(\mathrm{BR}(x')\) when \(x\) and \(x'\) are far apart, the optimizer must play enough rounds to adjust the average history from one strategy to the other. Our key insight is to design an algorithm that only makes local changes—ensuring that successive queries remain close—thus circumventing the high cost associated with large shifts in the history.

    \item \textbf{Discontinuities in the Objective:}  
    One might naturally attempt to apply standard local-search methods to our problem since our algorithm operates via local steps. However, the optimization target is not globally continuous: the learner's best response can change abruptly as the optimizer's mixed strategy crosses the boundaries between different regions of \(\Delta(\mathcal{A})\). Within each such best-response region—which is always a polytope—the principal's utility is continuous, but the overall function is piecewise continuous. Our algorithm addresses this challenge by detecting the hyperplanes that separate these best-response polytopes and restricting the search to the regions that yield higher utility.

    \item \textbf{Searching across High-Dimensional Polytope Boundaries:}  
    Our algorithm operates locally within polytopes, shifting from one to another upon reaching a local optimum within the current region. Once an \(x \in \Delta(\mathcal{A})\) is found that locally maximizes the optimizer's utility in a best-response polytope, the algorithm must determine whether any neighboring polytopes offer a higher utility. Detecting all adjacent regions is challenging because these polytopes reside in \(\mathbb{R}^{|\mathcal{A}|-1}\), and in high dimensions some may have volumes that are exponentially small relative to the dimension—even within a small neighborhood of \(x\). We overcome this by sequentially detecting neighboring polytopes and, at each step, restricting the search to the intersection of all previously discovered regions. This gradual reduction in the search space's effective dimension avoids the pitfall of runtimes that scale inversely with the volume of the smallest polytope. In contrast to prior query-based algorithms for finding global Stackelberg equilibria—which may incur exponential runtimes due to this volume dependence—we demonstrate that focusing on local Stackelberg equilibria sidesteps this exponential barrier.
\squishend

\subsection{Related Work}  
While the Stackelberg equilibrium can be computed in polynomial time given full information about the game via a reduction to linear programming, learning the Stackelberg equilibrium from a best-response oracle requires additional assumptions. This challenge has been analyzed in general games~\citep{letchford2009learning} and in specific games such as security games~\citep{letchford2009learning,blum2014learning,peng2019learning,balcan2015commitment}, demand learning~\citep{kleinberg2003value,besbes2009dynamic} and strategic classification~\citep{dong2018strategic,chen2020learning,ahmadi2023fundamental}.

Learning the Stackelberg equilibrium through interactions with a learning agent that can (relatively) quickly forget has been studied by reducing the problem to a query-based algorithm with access to an approximate best-response oracle. This approach has been applied to interactions with non-myopic agents in auction design~\citep{amin2013learning,mohri2014optimal,liu2018learning,abernethy2019learning} and in general games~\citep{haghtalab2022learning} and to interactions with adaptively calibrated agents \citep{haghtalab2024calibrated}. In contrast, when interacting with mean-based learners that do not forget quickly, both exponential lower and upper bounds have been established on the iteration complexity of approximating the Stackelberg value~\citep{brown2024learning}. 

Beyond the Stackelberg value, alternative benchmarks have been explored. Prior work has investigated the problem of finding a sequence of actions for the optimizer that maximizes its utility when interacting with a learning agent—potentially achieving a higher utility than that obtained by playing the Stackelberg equilibrium. This has been studied in both general game settings and in specific domains such as auction design, contract design, and information design~\citep{braverman2018selling, deng2019strategizing, mansour2022strategizing, cai2023selling, rubinstein2024strategizing, guruganesh2024contracting, lin2024persuading}. However, in general games, even when the optimizer has full knowledge of the learner's utility function, the optimization task is known to be NP-hard \citep{assos2024maximizing, assos2024computational}.  

Additional related work includes the impossibility of approximating the Stackelberg value when the learner is strategic or noisy~\citep{ananthakrishnan2024knowledge,donahue2024impact}; designing learners that are Pareto-optimal against strategic agents~\citep{arunachaleswaran2024pareto}; learning with Bayesian knowledge about the opponent~\citep{arunachaleswaran2024learning}; and computing the Stackelberg equilibrium in continuous games \citep{fiez2020implicit, maheshwari2023convergent, brown2024online}.

\newcommand{\optimizer}{principal\xspace}
\newcommand{\learner}{agent\xspace}
\newcommand{\principal}{principal\xspace}
\newcommand{\agent}{agent\xspace}
\newcommand{\aspaceopt}{\cA}
\newcommand{\aspacelearn}{\cB}
\newcommand{\sspaceopt}{\Delta(\aspacep)}
\newcommand{\sspacelearn}{\Delta(\aspacea)}
\newcommand{\aopt}{a}
\newcommand{\alearn}{b}
\newcommand{\sopt}{x}
\newcommand{\slearn}{y}
\newcommand{\uopt}{U_1}
\newcommand{\ulearn}{U_2}
\newcommand{\utilp}{U_1}
\newcommand{\utila}{U_2}
\newcommand{\utilpmatrix}{\boldsymbol{U_1}}
\newcommand{\utilamatrix}{\boldsymbol{U_2}}
\newcommand{\barutilamatrix}{\boldsymbol{\overline{U}_2}}
\newcommand{\noisematrix}{\boldsymbol{W}}
\newcommand{\noise}{W}
\newcommand{\Normal}{\mathcal{N}}
\newcommand{\aspacep}{\cA}
\newcommand{\aspacea}{\cB}
\newcommand{\sspacep}{\Delta(\aspacep)}
\newcommand{\sspacea}{\cY}
\newcommand{\ap}{a}
\renewcommand{\aa}{b}
\renewcommand{\sp}{x}
\newcommand{\sa}{y}
\newcommand{\up}{U_1}
\newcommand{\ua}{U_2}
\newcommand{\barua}{\overline{U}_2}
\newcommand{\sqsubmatrix}{\mathcal{S}}

\newcommand{\dimlearn}{n}
\newcommand{\dimopt}{m}
\newcommand{\bropt}{\BR_1}
\newcommand{\brlearn}{\BR_2}

\newcommand{\dimp}{n}
\newcommand{\dima}{m}
\newcommand{\brp}{\BR_1}
\newcommand{\bra}{\BR_2}

\renewcommand{\polytope}{\mathrm{P}}
\newcommand{\hatpolytope}{\hat{\polytope}}
\newcommand{\surroundingP}{\mathcal{P}}
\newcommand{\distance}{\mathrm{dist}}

\newcommand{\sigmalb}{\underline{\sigma}}

\newcommand{\rinf}{R_{\min}}
\newcommand{\rone}{R_1}
\newcommand{\rtwo}{R_2}
\newcommand{\rpoly}{\rho}

\newcommand{\bnew}{b_{\mathrm{new}}}
\newcommand{\hatsubspace}{\hat{\mathcal{S}}}

\section{Model}
In this paper, we consider a 2-player game between a \emph{principal} and an \emph{agent}.
Let $\aspacep$ be the action space of the principal and $\aspacea$ be the action space of the agent.
We assume that both action spaces are finite with size $|\aspaceopt| = \dimopt$ and $|\aspacelearn| = \dimlearn$, respectively. 

For a pair of pure strategies $(a,b)\in\aspacep\times\aspacea$, we use $\utilp(a,b)$ and $\utila(a,b)$ to denote the utility of the principal and the agent, where we assume that all utilities are between $[0,1]$. These utility functions can also be interpreted as matrices $\utilp,\utila\in[0,1]^{m\times n}$.

When both players employ mixed strategies, i.e. distributions $\x\in\Delta(\aspacep)$ and $\y\in\Delta(\aspacea)$ over their action spaces, their expected utilities are given by
\begin{align*}
    U_i(\x,\y)=\E_{a\sim\x,b\sim\y}[U_i(a,b)]=\x^{\transpose}U_i\ \y,
    \quad i\in\{1,2\}.
\end{align*}

\paragraph{Repeated Games.}
We consider repeated interactions between the principal and the agent, where the stage game $(\utilp,\utila)$ is repeated for $T$ rounds.
At the beginning of the interactions, both players know their own utility matrix $U_i$ but do not know the utility matrix $U_{-i}$ of their opponent. 

At each round $t \in [T]$, the principal and agent simultaneously select strategies $\x^{(t)}\in\Delta(\aspacep)$ and $\y^{(t)}\in\Delta(\aspacea)$ respectively. Principal observes $\y^{(t)}$, and gains utility $\utilp(\x^{(t)},\y^{(t)})$. Similarly, the agent observes  $\x^{(t)}$, and gains utility $\utila(\x^{(t)},\y^{(t)})$.

\subsection{Principal's Benchmarks}
To measure the principal's performance in repeated games, many previous works (e.g.~\citep{haghtalab2022learning,haghtalab2024calibrated,deng2019strategizing,blum2014learning}) have focused on the Stackelberg value of the stage game, which we introduce below.
\paragraph{The Stackelberg Value.}
In the stage game $(\utilp,\utila)$, the Stackelberg value is a benchmark for the principal's optimal utility when she has full knowledge of the agent's utility function and the ability to commit to a strategy.
Formally, it is defined as the solution to the following optimization problem:
\[
\stackval\triangleq
\max_{\x^\star\in\Delta(\aspacep)}
\max_{y^\star\in\BR(\x^\star)}
\utilp(\x^\star,y^\star),
\]
where  $\BR:\sspaceopt \to 2^\aspacelearn$ is the agent's best-response function that maps from a principal's mixed strategy to a set of pure strategies in $\aspacea$ that maximizes the agent utility. Specifically, for all $\x\in\Delta(\aspacep)$,
\[
\BR(\x) \triangleq \argmax_{\alearn \in \aspacelearn} \ulearn(\x,\slearn) = \{\alearn \in \aspacelearn :
\ulearn(\sopt, \alearn) \geq \ulearn(x, y) \text{ for all } y \in \sspacelearn \}.
\]

The Stackelberg value represents the global optimal utility that can be achieved against rational agents. However, computing it efficiently is intractable against mean-based agent, as shown by~\citet{brown2024learning}. In this paper, we study the \emph{local Stackelberg equilibria}, which we show can be achieved efficiently.

\paragraph{Local Stackelberg Equilibria (LSE).}
A local Stackelberg strategy is one where no small local deviation can significantly improve the principal's utility when the agent best responds.
This definition serves as
the discrete analogue of the differential Stackelberg equilibria studied in~\citep{fiez2020implicit}.

\begin{definition}[$(\eps,\delta)$-Approximate Local Stackelberg Strategy]
    A principal's strategy $\x\in\Delta(\aspacep)$ is an $(\eps,\delta)$-Approximate Local Stackelberg Strategy if \[
   \forall \x'\in\ball_1(\x;\delta),\quad 
   \sup_{y'\in\BR(\x')}\utilp(\x', y') \leq \sup_{y\in\BR(\x)}\utilp(\x, y)+\eps\delta,
    \]
    where $\ball_1(\x,\delta)$ denote the $\ell_1$ ball of radius $\delta$ around $\x$, i.e., the set of strategies with $\|\x'-\x\|_1\le\delta$.
\end{definition}
We remark that in settings with smooth utility functions, approximate local optima are often characterized by small gradients. However, when the principal's utility function is non-continuous due to the agent's best-response behavior, a gradient-based characterization is no longer appropriate. Instead, our definition provides a discrete analogue that captures the same intuition—ensuring that small perturbations in the principal's strategy do not lead to significantly higher payoffs. 

We include a comparison of the local Stackelberg benchmark to other benchmarks in \Cref{app:benchmarks}. In general, the local Stackelberg is a weaker benchmark compared to the global Stackelberg, albeit a tractable one. However, there are special cases, where the local Stackelberg is equivalent to the the global Stackelberg in terms of the principal's utility. This is the case in a broad and commonly studied class of Stackelberg games --- \emph{Stackelberg security games} as we show in \Cref{prop:SSG}.

\subsection{Agent's algorithm}
In this paper, we assume that the agent selects their actions according to a \emph{mean-based} learning algorithm, defined as follows:
\begin{definition}[Mean-based Algorithm~\citep{braverman2018selling}]\label{def:mean-based}
For each of the agent's action $y\in\aspacea$, let $\bar{\utila}^{(t-1)}(y)=\frac{1}{t-1}\sum_{s =1}^{t-1} \utila(\x^{(s)},y)$ denote the cumulative reward of action $y$ up to round $t$. 
The agent's learning algorithm is called \emph{mean-based} if there exists a sequence $(\mu_t)_{t\ge1}$ with $\mu_t = o_t(1)$ such that, at any round $t$, whenever an action $i\in\aspacea$ has
$
\bar{\utila}^{(t-1)}(i) < \max_{y\in \aspacea}\overline{\utila}^{(t-1)}(y)- \mu_t$,
then the probability of selecting $y_t=i$ is at most $\mu_t$.

\end{definition}

If an algorithm is mean-based with $\mu_t\equiv0$, then it reduces to the \emph{fictitious play} algorithm. The fictitious play algorithm satisfies $y^{(t)}\in\BR(\avgx^{(t-1)})$ per round, where $\avgx^{(t-1)} = \frac{1}{t-1}\sum_{s=1}^{t-1} \sopt^{(s)}$ is the principal's average strategy up to round $t$. 

We focus on mean-based agents since the class of mean-based algorithms includes many widely-used learning algorithms, such as Multiplicative Weights Update, Follow the Regularized/Perturbed Leader and $\eps$-greedy~\citep{braverman2018selling}. Moreover,
mean-based agents exhibit a property of slowly forgetting past interactions, as they update their strategy based on the average of the principal's past strategies. This introduces technical challenges that differ significantly from interactions with myopic best-responding agents who make decisions only based on recent rounds without any memory.

\subsection{Geometric Interpretations of the Principal's Optimization Problem}
In this section, we introduce some additional notations and revisit the geometric properties of the principal's optimization problem that will be useful in our algorithms.

For each of the agent's actions $b\in\aspacea$, define the \emph{best response polytope} to action $b$ as the set of principal's mixed strategies that induce best response $b$, which we denote with $\polytope_b\triangleq \left\{\x\in\Delta(\aspacep)\mid 
b\in\BR(\x)\right\}.$
These subsets are referred to as polytopes 
because they are characterized by the intersection of halfspaces:
\[
\polytope_b=\left\{\x\in\Delta(\aspacep)\mid \forall b'\in \aspacea,\ 
\utila(\x,b)-\utila(\x,b')=
\ip{\util_b-\util_{b'}}{\x}\ge0\right\},
\]
where $\util_b\in\br^{m}$ represents the agent's utility vector conditioned on action $b$, i.e., $\util_b=\utila(\cdot,b)$.
We will also use $\h_{b,b'}$ to denote the hyperplane that separates polytopes $\polytope_b$ and $\polytope_{b'}$, i.e., $\h_{b,b'}\triangleq\util_b-\util_{b'}$.

\paragraph{Approximate Best Response Oracle.} A key property of mean-based agents is that it enables us to implement approximately best response oracles. More specifically, we construct $\broracle$ in \Cref{sec:br_oracle}, which satisfies that whenever the principal's average strategy $\bar{\x}^{(t)}$ lies robustly within a best-response polytope $\polytope_y$, the oracle $\broracle(\bar{\x}^{(t)})$ will return the best-response strategy $y=\BR(\bar{\x}^{(t)})$ with high probability. 

\paragraph{Principal's Optimization Problem.} Note that under full information about the polytope partition, the principal's Stackelberg value can be equivalently described as the optimal solution to the following piece-wise linear function:
$
\stackval=\max_{b\in\aspacea}
\max_{\x\in\polytope_b} 
\utilp(\x,b),
$
which involves maximization over all the polytopes $\polytope_b$. On the other hand, in \emph{local Stackelberg equilibria}, a strategy $\x$ is locally optimal if no surrounding polytopes achieve a higher principal utility. We define the notion of (approximately) surrounding polytopes below.

\paragraph{Surrounding Polytopes.}
For a principal's strategy $\x\in\Delta(\aspacep)$, let $\surroundingP(\x)$ denote the subset of polytopes that contains (surrounds) $\x$: $\surroundingP(\x)\triangleq\{\polytope_b\mid \x\in\polytope_b\}$. Note that $\surroundingP(\x)$ equivalently contains all polytopes $\polytope_b$ for which $b\in\BR(\x)$ is a best response.

For a radius $\eps>0$, let $\surroundingP_\eps(\x)$ be the polytopes that are within $\ell_2$ distance $\eps$ to $\x$: \[\surroundingP_\eps(\x)\triangleq\{\polytope_b\mid \distance_2(\x,\polytope_b)\le \eps\}.
\tag{$\eps$-Surrounding Polytopes}\] Here, the distance between vector $\x$ and set $\polytope_b$ is defined as 
$\distance_2(\x,\polytope_b)\triangleq \min_{\x'\in\polytope_b}\|\x-\x'\|_2$.

\subsection{Our Assumptions}
\label{sec:assumptions}
In this section, we list the assumptions we make to derive our results. We provide a more detailed discussion in \Cref{app:assumptions}.

Our first assumption is a standard assumption in previous work on approximating the Stackelberg value \citep{blum2014learning,letchford2009learning,haghtalab2022learning,haghtalab2024calibrated}. 

\begin{assumption}[Distance from Polytope Boundaries]
    \label[assumption]{assump:far-from-boundary}
        For every polytope $\polytope_b$, there exists some strategy $\x\in\polytope_b$ such that all coordinates of $\x$ are lower bounded by $\rinf$, i.e., $\x\ge \rinf\cdot\vecone$.
    \end{assumption}
    \Cref{assump:far-from-boundary} is an arguably weaker condition than the assumption made in previous works as discussed in \Cref{remark:far-from-boundary}. One reason is previous works~\citep{blum2014learning,letchford2009learning,haghtalab2022learning,haghtalab2024calibrated} usually incur a dependency on the volume of the ball, which can be exponential in the dimension (i.e. the number of principal's actions). In contrast, our bound only has a polynomial dependence on $1/\rinf$.

The next assumption is that the polytopes and hyperplanes are sufficiently separated from each other. In \Cref{app:assumptions}, we show how this assumption can be implied by a lower bound on the minimum singular value of constraint matrices defining the polytopes, which we denote by $\sigmalb$. 
Here we state it informally as follows:

\begin{assumption}[Informal version of \Cref{lemma:polytopes-far}: Polytopes/Hyperplanes are Far Apart]
    \label[assumption]{inf_assumption:polytopes-far}
    \leavevmode
    \begin{enumerate}
        \item For any $\x \in \Delta(\aspacep)$, there are at most $m$ polytopes that have distance at most $R_1$ to $\x$.
        \item 
        For all $b\in\aspacea$ and $\x\in\polytope_b$, there are at most $m-1$ hyperplanes from $\polytope_b$ that $\x$ can be close to.
    \end{enumerate}
\end{assumption}

The second assumption above is satisfied with high probability under the smoothed analysis framework i.e., when the utility functions are perturbed by Gaussian noise. We show this in \Cref{sec:smoothed-analysis}.

Our final assumption follows from \cite{haghtalab2022learning}, and states that the normalized vectors defining the agent's utilities are sufficiently different, which ensures that the mean-based property can be used to approximate the best response for principal strategies sufficiently far from polytope boundaries.   

\begin{assumption}[Regularity of agent's utility~\citep{haghtalab2022learning}]  
\label[assumption]{assump:agent-utility-different}
For each of the agent's action $y\in\aspacea$ and principal's action $x\in\aspacep$, let $v^{y}(x)\triangleq \utila(x,y)-\frac{1}{|\aspacep|}\sum_{x'\in\aspacep}\utila(x',y)$, then the vectors $\{v^y\}_{y\in\aspacea}$ satisfies $\min_{y\neq y'}\|v^y-v^{y'}\|_2\ge\Delta$.
\end{assumption}

\section{Algorithm}
In this section, we present our main algorithm  (\Cref{alg:lse_alg}) that finds an approximate Local Stackelberg equilibrium. \Cref{alg:lse_alg} alternates between the following two main subroutines: 
\begin{itemize}[leftmargin=1em]
    \item $\optwithinsubspace$ (\Cref{alg:optimize-within-polytope}). At a high level, this subroutines finds an approximately optimal policy constrained to a given polytope.
    \item $\searchpolytope$ (\Cref{alg:search-for-polytope}). When the principal's search point is near a polytope boundary, this subroutine identifies all adjacent polytopes in the neighborhood.
\end{itemize}
 
Now we discuss the high-level ideas for the two subroutines respectively.

\textbf{(1) Optimizing within polytopes (More details in \Cref{sec:within-polytopes}).}
In this subroutine, the principal aims to compute an $\varepsilon \delta$-optimal strategy within the current polytope.
This problem would be a linear program if the polytope were fully known. However, because the polytope depends on the agent's private utilities (which are unknown to the principal), the principal must learn this structure through interactions. To do so, the principal maintains an \emph{approximate} version of the polytope, which is iteratively refined as constraints become known.

The approximation to the polytope is learned incrementally by adding new constraints as new polytopes are encountered. After each update of the constraints, the principal chooses the optimal strategy among strategies satisfying the current set of constraints. In doing so, the principal either improves utility or discovers a new constraint that refines the polytope approximation and prompts a re-optimization. (As shown in \Cref{fig:opt-within-polytope}.)

\textbf{(2) Searching for new polytopes (More details in \Cref{sec:search-polytopes}).}
The $\searchpolytope$ subrountine explores all polytopes within a certain radius around a given point. 
A key challenge in this search is that some surrounding polytopes may have volume that is exponentially small in the number of dimensions, requiring exponential in dimensions number of samples to identify through random sampling.
To efficiently discover these polytopes, $\searchpolytope$ employs an iterative dimension reduction approach. After a subset of polytopes are found, the algorithm restricts its search to the boundaries these polytopes. This restriction reduces the dimension of the search space and overcomes the challenge of searching in a high dimensional space. (See \Cref{fig:search-polytopes} for an illustration.)

\begin{algorithm}[t]
  \caption{Local Stackelberg Equilibrium}
  \label{alg:lse_alg}
  \Input{$x_{\text{start}}, \varepsilon, \delta, \alpha, \gamma$}
  \Output{An $(\varepsilon, \delta)$-approximate Local Stackelberg equilibrium}

  \tcp{Iterating through the best-response polytopes}
  Burn-in period: play arbitrary strategy for $t_0$ rounds until mean-based parameter $\mu_{t_0} \le \alpha / 2$.\;

  \For{$i = 1, \ldots, n$}{
    Current average strategy: $\overline{\mathbf{x}}$.\;
    Current agent best-response: $b_i \gets \broracle(\overline{\mathbf{x}})$.\;
    Update average strategy to $\mathbf{x}^*_i \gets \optwithinsubspace(\BR(\overline{\mathbf{x}}), \varepsilon, \delta, \alpha, \gamma)$.\;
    Move average strategy $\mathbf{x}^*_i$ to be $\delta$-close to the boundary.\;
    Find all surrounding polytopes using $\searchpolytope(\mathbf{x}^*_i)$.\;

    \If{there exists a surrounding polytope $\polytope_b$ with
        $\uopt(\mathbf{x}^*_i, b) \ge \uopt(\mathbf{x}^*_i, b_i) + \varepsilon_2$}{
        Step into polytope $b$ and continue to the next iteration.\;
    }
    \Else{
        \KwRet{$(\mathbf{x}^*_i, \BR(\mathbf{x}^*_i))$ is an $(\varepsilon,\delta)$-approximate LSE.}
    }
  }
\end{algorithm}

\begin{theorem}[Main theorem]\label{thm:main}
Under \Cref{assump:singular-value,assump:far-from-boundary}, with high probability, 
\Cref{alg:lse_alg} finds an $(\varepsilon, \delta)$-approximate Local Stackelberg equilibrium within the following number of iterations: \[
\textstyle
O\left(\text{poly}\left(m, n, \frac 1 \rinf, \log \frac 1 \sigmalb\right) \cdot \exp\left(\frac1 {\varepsilon \delta}\right)\right)\]
\end{theorem}

Although the number of iterations needed for \Cref{alg:lse_alg} to approximate the local Stackelberg equilibrium is exponential in $1/(\varepsilon \delta)$, we show in the following theorem that this exponential dependence is unavoidable. The proof of \Cref{thm:lower_bound} is deferred to \Cref{proof:lower_bound}.

\begin{proof}[Proof sketch of \cref{thm:main}.]
In this proof sketch we will illustrate how we bound the number of rounds the algorithm takes. In the next sections we will analyze the subroutines of the algorithm in more detail and analyze their correctness. The correctness of the algorithm will follow from the correctness of these subroutines.

\newcommand{\alphaone}{\delta}
\newcommand{\alphatwo}{\frac{\sigmalb \rtwo}{m^3}}
\newcommand{\alphathree}{\frac{\gamma \sigmalb^2}{n^2 m^7 \log m}}
\newcommand{\alphafour}{\frac{\varepsilon \sigmalb}{m}}

    The main argument to bound the number of rounds our algorithm takes to find an $(\varepsilon, \delta)$-approximate local Stackelberg strategy is that in each round where we improve utility, the principal's utility improves by at least some minimum amount.
    The amount of utility improvement has to necessarily depend on the round $t$ since the maximum possible change in magnitude of utility at round $t$ is $1/t$, as shown in \Cref{remark:max-step-size}. This follows from how the average strategy of the principal changes with the choice of strategies in each round. (We provide details on how our algorithm chooses strategies to change the average strategy in \Cref{sec:query_average_strategies}.)
    We first suppose each improvement to the utility the algorithm makes has magnitude at least $\Delta / t$, and show that our algorithm terminates within a bounded number of rounds leveraging the fact that the total possible improvement of utility is bounded. We will prove a lower bound on $\Delta$ at the end of the proof.
    
    The argument is similar to the convergence analysis of algorithms such as gradient descent in continuous settings to local optima. However, a key challenge  in our setting is that not every step is an improvement due to the discontinuous (piecewise) nature of the principal's utility functions. To deal with this discountinuity, our algorithm also takes non-improvement steps (such as the steps for constructing polytope boundaries and  $\searchpolytope$).

    Suppose the set of improvement rounds is $T_I$ and the set of other rounds is $T_N$.
    We establish an upper bound on $|T_N|$ which allows us to still upper bound the iteration complexity of approximating the local Stackelberg strategy since each non-improvement step decreases utility by at most $\frac{1}{t}$.
    The total utility improvement is at least
    \newcommand{\impnum}{|T_{\text{I}}|}
    \newcommand{\nonnum}{|T_{\text{N}}|}
    \newcommand{\imp}{T_{\text{I}}}
    \newcommand{\nonimp}{T_{\text{N}}}
    \begin{align*}
    \sum_{t \in \imp} \frac{\varepsilon \delta}{t} - \sum_{t \in \nonimp} \frac{1}{t}
    \ge 
    \sum_{t = \nonnum}^{\nonnum + \impnum} \frac{\Delta}{t} - \sum_{t =1}^{\nonnum} \frac{1}{t} 
        \gtrsim \Delta \log \frac{\nonnum + \impnum}{\nonnum^2 }.
    \end{align*}
    Since utilities are bounded in $[0,1]$,
    \begin{align*}
        \textstyle
        \Longrightarrow \Delta \log \frac{\nonnum + \impnum}{\nonnum^2} &\leq 1, 
        \quad \quad \nonnum + \impnum \leq \nonnum^2 + \exp \left (\frac 1 {\Delta} \right ).
    \end{align*}

    To complete the bound on the total number of rounds, it suffices to find a lower bound on the minimum utility improvement quantity $\Delta$ and an upper bound on the number of non-improving rounds $\nonnum$.
    We will show in \Cref{sec:within-polytopes} that $\Delta \geq \varepsilon \delta$, in the analysis of $\optwithinsubspace$ (the subroutine where utility improvement steps are made). 

    Non-improvement include the burn-in period and the rounds to construct boundary hyperplanes when an intended improvement step takes us outside the polytope. The burn-in period $t_0$ is polynomial in $1/\alpha$ as long as mean-based parameter $\mu_t$ decays at least polynomially in $t$ (as is the case for canonical mean-based algorithms such as Multiplicative weghts, Follow the Perturbed Leader).
    
    The rounds to construct boundary hyperplanes occurs at most $n^2$ times, once to construct easy hyperplane. And each round includes steps to get close to the boundary via $\mathsf{BinarySearch}$ and generating search vectors to explore nearby polytopes in $\searchpolytope$. The number of such steps $\nonnum$ is analyzed in the analysis of the sub-routines and will turn out to be $\mathrm{poly}(m,n) \exp(1/(\varepsilon \delta))$. This results in the bound of the theorem.
    \end{proof}

\newcommand{\optwithinpolytopealgo}{%
    \begin{algo}
    {\caption{$\optwithinsubspace$}}
    \label[algo]{alg:opt-within-polytopes}                 %
      {%
        \DontPrintSemicolon
        \SetKwInOut{Input}{Input} \SetKwInOut{Output}{Output}
        \Input{Current polytope $b$, starting point $\x_{\text{start}}$, approximation parameter $\alpha$}
        
        $\hat{\mathcal P}_b \gets \Delta(\mathcal A)$\;  %
        \While{$\hat{\mathcal P}_b$ is not empty}{%
          $\x_{\text{target}} \gets \argmax_{\x \in \hat{\mathcal P}_b} U_1(\x, b)$\;
          \If{$U_1(\x_{\text{target}}, b) < U_1(\x_{\text{start}}, b) + \varepsilon \delta$}{%
            \KwRet $\x_{\text{start}}$\;
          }
          
          $\bar{\x}^{(t)} \gets \x_{\text{target}}$\;  \tcp*{improvement step}
          \If{$y^{(t)}=b'\neq b$}{%
            $(\x,\x') \gets \binarysearch\,$ between $\bar{\x}^{(t-1)}\in\polytope_b$ and $\bar{\x}^{(t)}\in\polytope_{b'}$ with accuracy $\alpha$\;
            Use $(\x,\x')$ to find approximate hyperplane $\hat{\h}_{b,b'}$ by calling $\searchpolytope$\;
            $\hat{\mathcal P}_{b} \gets \hat{\mathcal P}_b \cap \{\x: \hat\h_{b,b'}^{\transpose}\x \ge \alpha\}$\;
            $\x_{\text{start}} \gets \text{projection of }\x_{\text{start}}\text{ onto }S$\;
          }
        }
      }%
  \end{algo}%
}

\newcommand{\searchpolytopealgo}{
\begin{algo}
\caption{$\searchpolytope$}\label[algo]{alg:search-for-polytope}
    \KwIn{Starting point $\x^\star$, accuracy level $\alpha$}
    $\hatsubspace \gets \{\x: \vecone^{\transpose} \x = 1 \}$ (the constrained search space)\;
    $L \gets \{\}$ (the set of hyperplanes and polytopes discovered so far)\;
    \While{$\hatsubspace$ { is not empty}}{
    $\h,b \gets \mathsf{FindAHyperplane}(\x^\star, \hatsubspace, \alpha)$\;
    \If{$\h,b$ is None}{Return $L$}
    $\hatsubspace \gets \hatsubspace \cap \{x: \ip{\h}{\x} = \alpha\}$\;
    $L \gets L \cup \{(\h,b)\}$ \;
    $\x^\star \gets \text{projection of $\x^\star$ onto $\hatsubspace$}$\;
}
\end{algo}
}

\begin{algorithm}[t]
  \caption{OptimizeWithinPolytope}
  \label{alg:optimize-within-polytope}
  \Input{Current polytope $b$, starting point $\mathbf{x}_{\text{start}}$, approximation parameter $\alpha$}

  $\hat{\mathcal{P}}_{b} \gets \Delta(\mathcal{A}_1)$\;
  \While{$\hat{\mathcal{P}}_{b} \neq \varnothing$}{
    $\mathbf{x}_{\text{target}} \gets \arg\max_{\mathbf{x}\in\hat{\mathcal{P}}_{b}} U_1(\mathbf{x},b)$\;
    \If{$U_1(\mathbf{x}_{\text{target}},b) < U_1(\mathbf{x}_{\text{start}},b)+\varepsilon\delta$}{
      \KwRet{$\mathbf{x}_{\text{start}}$}\;
    }
    $\mathbf{x},\mathbf{x}' \gets \textsc{BinarySearch}$ between
    $\bar{\mathbf{x}}^{(t-1)}\!\in\!\mathcal{P}_b$ and $\bar{\mathbf{x}}^{(t)}\!\in\!\mathcal{P}_{b'}$ with accuracy $\alpha$\;
    Use $\mathbf{x},\mathbf{x}'$ to find $\hat{\mathbf{h}}_{b,b'}$ via $\searchpolytope$\;
    $\hat{\mathcal{P}}_{b}\gets \hat{\mathcal{P}}_{b}\cap\{\mathbf{x}:\hat{\mathbf{h}}_{b,b'}^{\top}\mathbf{x}\ge \alpha\}$\;
    $\mathbf{x}_{\text{start}}\gets \text{projection of $\mathbf{x}_{\text{start}}$ onto $S$}$\;
  }
\end{algorithm}

\begin{algorithm}[t]
  \caption{SearchForPolytopes}
  \label{alg:search-for-polytope}
  \Input{Starting point $\mathbf{x}^{\star}$, accuracy level $\alpha$}
  \Output{List $L$ of discovered hyperplanes/polytopes}

  $\hat{S}\gets\{\mathbf{x}:\mathbf{1}^\top\mathbf{x}=1\}$ \tcp*{Constrained search space}
  $L\gets\varnothing$ \tcp*{Discovered hyperplanes/polytopes}
  \While{$\hat{S}\neq\varnothing$}{
    $(\mathbf{h},b)\gets \textsc{FindAHyperplane}(\mathbf{x}^{\star},\hat{S},\alpha)$\;
    \If{$(\mathbf{h},b)=\textsc{None}$}{\KwRet{$L$}}
    $\hat{S}\gets \hat{S}\cap\{\mathbf{x}:\langle \mathbf{h},\mathbf{x}\rangle=\alpha\}$\;
    $L\gets L\cup\{(\mathbf{h},b)\}$\;
    $\mathbf{x}^{\star}\gets \Pi_{\hat{S}}(\mathbf{x}^{\star})$ \tcp*{Project onto $\hat{S}$}
  }
\end{algorithm}

\newcommand{\searchpolytopefigure}{
    \begin{tikzpicture}[scale=1, every node/.style={font=\footnotesize}]
    
    \coordinate (O) at (0,0);
    
    \draw[thick] (-1,-1) -- (O) -- (1,-1);
    \draw[thick] (O) -- (0,1.5);
    
    \node[left] at (-1,0.1) {\(\polytope_{b_1}\)};
    \node[right] at (0.5,0.8) {\(\polytope_{b_2}\)};
    \node[below] at (0,-1) {\(\polytope_{b_3}\)};
    
    \draw[dashed, red, thick] (-0.2,1.5) -- (0.04,-1.2);
    \node[left, red, align=center] at (-0.2,1) {Approx. \\hyperplane\\ \(\hat{\h}_{b_1,b_2}\)};
    
    \draw[->, ultra thick, blue] (-0.05,0) -- (-0.12,0.5);
    \node[right, align=center, blue] at  (0,0) {Random search\\ along $\hat{\h}_{b_1,b_2}$};
    \draw[->, ultra thick, blue] (-0.05,0) -- (0,-0.5);

    \coordinate (R) at (-0.05,0);
    \fill (R) circle (2pt);
    \node[left] at (R) {$\x_0$};

    \coordinate (N) at (0,-0.7);
    \fill (N) circle (2pt);
    \node[right, align=center] at (N) {$\avgx^{(t)}$};
    
    \end{tikzpicture}
}

\subsection{Analysis of $\optwithinsubspace$}
\label{sec:within-polytopes}
In this section we will analyze the correctness of 
the subroutine $\optwithinsubspace$. Before we delve into the analysis of $\optwithinsubspace$, let us describe its implementation in some more detail. 

When optimizing within a polytope $\polytope_b$ corresponding to action $b \in \cB$, $\optwithinsubspace$ maintains an estimate polytope $\hatpolytope_b$ of the true polytope $\polytope_b$. $\hatpolytope_b$ starts of being the space of all strategies and gets refined as the principal learns more about $\polytope_b$. The algorithm alters between \emph{improvement} steps and \emph{searching} steps:

\usetikzlibrary{calc}

\newcommand{\optwithinpolytopefigure}{
    \begin{tikzpicture}[scale=0.7, every node/.style={font=\footnotesize}]
    
      \coordinate (A) at (0,0);
      \coordinate (B) at (4,0);
      \coordinate (C) at (2,3);
      
      \draw[thick] (A) -- (B) -- (C) -- cycle;
      \node at (2, -0.4) {Polytope $\polytope_b$};
      
      \coordinate (P) at (2.6,0.8);
      \fill (P) circle (2pt);
      \node[below left] at (P) {$\avgx^{(t-1)}$};
      
      \coordinate (R) at (3.3,2.3);
      \fill (R) circle (2pt);
      \node[right] at (R) {$\avgx^{(t)}$};
      
      \draw[->, ultra thick, blue] (P) -- (R) node[above right, align=center] {previous improvement \\direction $\util_b$};

      \coordinate (Bprime) at ($(B)-(0.05,0.05)$);
      \coordinate (Cprime) at ($(C)-(0.5,0)$);
      
      \draw[dashed, red, thick] (Bprime) -- (Cprime);
      \node[below right, red] at (Bprime) {Approx. boundary};
      \coordinate (Q) at (2.85,1.3);
      \fill (Q) circle (2pt);
      \draw[->, ultra thick, blue] (Q) -- ($(Q) + 0.3*(Cprime) - 0.3*(Bprime)$) node[left,align=center] {new improvement \\direction};
    
    \end{tikzpicture}
}

\squishlist
    \item In \emph{improvement} steps, we move the average strategy $\avgx^{(t)}$ toward the optimizer $\z=\argmax_{\x\in\hatpolytope_b}\utilp(\x,b)$, which is the optimizer in $\hatpolytope_b$. We move toward $z$ only when utility improves significantly. This guarantees termination via the termination condition \[
    \utilp(\z,b)\le
    \utilp(\avgx^{(t)},b)+\eps\delta.
    \tag{Termination condition}
    \]
    
    $z$ might not actually be in $\polytope_b$ and may lead to an average strategy $\avgx^{(t)}$ outside of $\polytope_b$ that can be detected by $y^{(t)} \neq b$ as shown in \Cref{fig:opt-within-polytope}. When this happens, we switch to \emph{searching} steps and identify a new direction of improvement if one exists as shown in \Cref{fig:opt-within-polytope}. 
    
    \item We enter a phase of \emph{searching} steps with two consecutive points $\avgx^{(t-1)}$ and $\avgx^{(t)}$ that lie in opposite sides of some boundary $(\h)$ of $\polytope_b$. We approximate $\h$ by $\hat{\h}$ satisfying $\|\hat{\h}-\h\|_2\le\alpha$.

    We do this by backtracking from $\avgx^{(t)}$ to $\avgx^{(t-1)}$ and finding a point in between that is $\alpha$ close to the boundary. We search the polytopes surrounding this point via $\searchpolytope$ to construct $\hat{\h}$. 
    
\squishend

This implementation of $\optwithinsubspace$ approximates the optimal strategy within the polytope of search as stated below.

\begin{proposition}[Correctness of $\optwithinsubspace$]\label[proposition]{prop:opt_within_polytope}
    \Cref{alg:optimize-within-polytope} takes an \agent-action $b \in \cB$ and a starting point in the best-response polytope $\polytope_b$,
    and finds a strategy $\x^\star_b \in \polytope_b$ that achieves $\eps\delta$-optimal principal's utility in the polytope, i.e.,
    $\up(\x^\star_b, b) \geq \max_{\x_b\in\polytope_b} \up(\x_b, b) - \eps \delta$.
\end{proposition}
\begin{proof}[Proof sketch]
    The correctness of $\optwithinsubspace$ is due to the way  we construct the estimated polytope $\hatpolytope_b$. The first property is that $\hatpolytope_b$ is defined by approximations to a subset of the constraints defining $\polytope_b$. As a result, if no strategy in $\hatpolytope_b$ significantly improves utility (over $\varepsilon \delta$ amount), then there is also no strategy in $\hatpolytope_b$ improving utility beyond $\varepsilon \delta$. 
    The second property is the accuracy to which we approximate the constraints of $\polytope_b$ that we discover. We achieve this accuracy by using $\binarysearch$ to get close to the boundary of $\polytope_b$ and using $\searchpolytope$ to discover all boundaries  nearby. $\binarysearch$ also ensure that we don't get too close to the boundary to allow us to remain within the interior of the polytope where the best-response oracle is able to find the best-response for mean-based learners. 
\end{proof}

Beyond the correctness of $\optwithinsubspace$, there are two other important properties that we discuss below. These properties are regarding the minimum improvement in every improvement step and the number of non-improvement steps. Both of these quantities are important for bounding the total number of rounds the main algorithm uses to find the approximate local Stackelberg equilibrium.

In each improvement step, utility increases by at least $\varepsilon \delta / t \log t$ as shown in \Cref{lemma:improvement-steps}. This is due to the choice of termination condition which implies that each improvement occurs in a direction with at least $\varepsilon \delta$ utility improvement possible. 

All non-improvement steps occur in the searching phase are the rounds in sub-routines $\binarysearch$ and $\searchpolytope$. These steps include back-tracking toward previous average strategies. This is made efficient by our algorithm's restriction of selecting strategies in $\Delta(\cA)$ that are at least at $\ell_1$ distance of $\gamma$ away from the boundary of the simplex. Given this restriction, changing the average strategy at round $t$ from  $\bar{\x}$ to $\bar{\x}'$ can be done in $\|\bar{\x} - \bar{\x}'\|_1 t / \gamma$ rounds.

\subsection{Analyzing $\searchpolytope$ Searching for polytopes} 
\label{sec:search-polytopes}

\begin{figure}[!htbp]
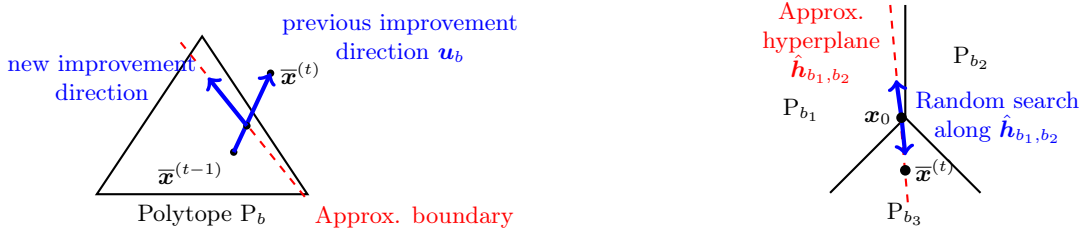

  \centering
  \begin{subfigure}[b]{0.47\textwidth}
    \centering
    \optwithinpolytopefigure
    \caption{Optimization within polytopes. The figure illustrates how improvement directions are updated in $\optwithinsubspace$ after adding new constraints. If $\avgx^{(t)}$ falls outside of the true polytope $\polytope_b$, we first find an approximate boundary (the dashed line), and update the improvement direction subject to the new constraint.}
    \label{fig:opt-within-polytope}
  \end{subfigure}
  \hfill
  \begin{subfigure}[b]{0.47\textwidth}
    \centering
    \searchpolytopefigure
    \caption{Search for polytopes. The figure illustrates the $\searchpolytope$ procedure. $\polytope_{b_1}$ and $\polytope_{b_2}$ are the discovered polytopes. When performing random search along the approximate hyperplane $\hat{\h}_{b_1,b_2}$, the search point $\avgx^{(t)}$ falls in $\polytope_{b_3}$ with constant probability. When this happens, we discover a new polytope $\polytope_{b_3}$.}
    \label{fig:search-polytopes}
  \end{subfigure}
  \caption{Algorithmic components for computing local Stackelberg equilibria.}
  \label{fig:combined}
\end{figure}

The $\searchpolytope$ algorithm finds all the polytopes that are within a distance of at most $\rpoly$ from a given point $\x\in\Delta(\aspacep)$. For illustration, consider the example in \Cref{fig:search-polytopes}. If the input point $\x$ is within $\rpoly$ distance to the intersection $\x_0$, then $\searchpolytope$ would return all the surrounding polytopes $\polytope_{b_1},\polytope_{b_2},\polytope_{b_3}$. This guarantee is formally stated in the following theorem.

\begin{restatable}[Correctness of $\searchpolytope$]{theorem}{thmsearchcorrectness}
\label{thm:search_correctness}
    Starting from a point $\x^*$ in $\polytope_b$, for any $\alpha \in (0, o(R_2 \sigmalb / m^3))$ and $\rho < \alpha$, $\searchpolytope$ finds $\hat{\h}_{b, b'}$ such that $\|\hat{\h}_{b, b'} - h_{b,b'}\|_2 \leq \alpha$, for every $b' \in \mathcal{P}_{\rho}(\x^*)$.
\end{restatable}

\begin{remark}\label{remark:apply_search_polytopes}
    Our algorithm applies $\searchpolytope$ in two contexts. The first is to find separating hyperplanes in $\optwithinsubspace$ and the second is to find all polytopes within a $\delta$ radius of a point to certify it as a Local Stackelberg equilibrium or find evidence against this.

    In the first context, we require $\alpha$-approximation of the separating hyperplane. Since the guarantee of the theorem holds for $\rho < \alpha$, we need to get within $\alpha$ close to the boundary so that the separating hyperplane lies in the region that $\searchpolytope$ explores. In the second context, we apply the theorem for the choice of $\rho = \delta$.
\end{remark}

As previously mentioned, the main challenge that $\searchpolytope$ overcomes is searching all polytopes efficiently including ones with exponentially small volume are difficult to find through random search.  $\searchpolytope$ still performs random search, but within a restricted space of a lower dimension. The restricted space is the intersections of all the boundary hyperplanes discovered up to the point of the search. 

That is, suppose boundary hyperplanes $\h_j$ for $j \in J$ are encountered and are approximated as $\hat{\h}_j$ for each $j \in [J]$. Then in the next round of search, $\searchpolytope$ conducts a Gaussian random search within the subspace $\hatsubspace_J:=\{
    \x\in\Delta(\aspacep)\mid 
    \hat{\h}_J\x=\veczero
    \}, 
    \label{eq:subspace}$ which has dimension $m - |J|$ as a consequence of \Cref{assump:singular-value}.

\Cref{inf_assumption:polytopes-far} ensures that all the random directions generated for search do indeed fall in true surrounding polytopes and not in any other polytope. This is because by the assumption, all other polytopes are sufficiently far enough. Hence we don't falsely discover polytopes other than the surrounding polytopes.

\section{Lower bound}
Our upper bound on the iteration complexity to approximate an LSE (\Cref{thm:main}) had a polynomial dependence on the dimensions of the problem, but exponential dependence on $1/\varepsilon$, where $\varepsilon$ is the level of approximation. In this section, we show that this exponential dependence is unavoidable. To illustrate this, the following theorem provides a lower bound on the iteration complexity when interacting with a learner employing the fictitious play.
\begin{restatable}[Lower Bound]{theorem}{thmlowerbound}
\label{thm:lower_bound}
    Assume a repeated game between an agent, who employs Fictitious Play, and a principal who does not know the agent's utility. Let $n$ be the number of actions for the agent, and let $\epsilon \in (0,1/3)$. Assume that $n \ge 1/\epsilon^2$. Let $m$ be the number of actions for the principal and assume that $m \ge 1/\epsilon^5$. Then, there exists a distribution over games such that for any randomized algorithm used by the principal, the expected number of iterations required to find an $\epsilon$-approximate local Stackelberg equilibrium is at least $e^{C/\epsilon}$, where $C>0$ is a universal constant.

    Further, this lower bound holds under smoothed analysis, when each entry in the agent's utility matrix is perturbed by a small constant and when Assumption~\ref{assump:far-from-boundary} is satisfied.
\end{restatable}

We present a proof sketch of \Cref{thm:lower_bound} in \Cref{proof:lower_bound}. Below, we sketch the intuition behind the construction. We construct distribution over games between the principal and the agent such that each game has a unique local Stackelberg strategy which is also the global Stackelberg strategy.
Specifically, our construction partitions the simplex into a large best-response polytope in the middle, and a path of approximately $\ell\approx1/\epsilon$ small polytopes centered around vertices $v_1, \dots, v_\ell$. The utility function is designed so that the principal's utility increases monotonically along this path, ultimately reaching its maximum at the final vertex $v_\ell$, which corresponds to the unique local (and global) Stackelberg equilibrium. Since the principal has no prior knowledge, she has to make progress by making the historical average strategy $\bar{\x}^{(t)}$ (which the agent approximately best responds to) follow this path in sequence. However, because the average strategy $\bar{\x}^{(t)}$ can only move by $O(\frac1t)$ at round $t$, traversing a distance of $\frac1\epsilon$ requires at least $\exp(\Omega(\frac{1}{\epsilon}))$ rounds, which gives rise to the exponential lower bound.

\section{Smoothed analysis}
\label{sec:smoothed-analysis}

In this section, we show that the singular value assumption (\Cref{assump:singular-value}) holds with high probability when the agent's utility matrix $\utilamatrix$ is obtained by perturbing any given utility matrix (denote the original matrix with $\boldsymbol{\utilamatrix}$) with i.i.d. Gaussian noise.

More formally, let $\boldsymbol{\barua}\in[0,1]^{m\times n}$ be the initial agent utility where each entry $\utila(a,b)\in[0,1]$ represents the agent's utility under action pair $(a,b)\in\aspacep\times\aspacea$. 
We will perturb this utility matrix by adding an independent Gaussian noise $\noise(a,b)\sim\Normal(0,\sigma^2)$ to each entry $(a,b)$, i.e., \[
\forall (a,b)\in\aspacep\times\aspacea,\quad
\utila(a,b)=\barua(a,b)+\noise(a,b).\]
In matrix form, this can be written as
$\utilamatrix=\boldsymbol{\barua} + \boldsymbol{W}$, where $\boldsymbol{W}\simiid \Normal(0,\sigma^2)$ is the Gaussian perturbation matrix. Recall that as defined in \Cref{def:constraint-matrix}, $\augconstraints_b$ denotes the augmented constraint matrix for any agent action $b\in\aspacea$, and $\sqsubmatrix_m(\augconstraints_b)$ denotes the set of all $m\times m$ square sub-matrices of $\augconstraints_b$.

\begin{restatable}[Lower bound on the minimum singular value]{theorem}{thmsmoothedanalysis}
    \label{thm:smoothed-analysis}
    Let $\barutilamatrix\in[0,1]^{m\times n}$ be an arbitrary utility matrix of the agent, and let $\utilamatrix$ be a Gaussian perturbation of $\barutilamatrix$ with variance $\sigma^2$. Then the resulting augmented constraint matrices of the perturbed utility matrix satisfies that for 
    $
    \sigmalb=\Theta\left(
        \frac{\sigma\delta}{m^{\frac{5}{2}}2^n}
    \right),$
    \Cref{assump:singular-value} holds with probability at least $1-\delta$. 
\end{restatable}
\begin{remark}
    Although $\sigmalb$ has an exponential dependency on $m$, in our final bound (\Cref{thm:main}), the dependence is only through $\log(1/\sigmalb)$. As a result, this introduces only a logarithmic dependence on $m$ and a polynomial dependence on $n$.
\end{remark}

To establish \Cref{thm:smoothed-analysis} under smoothed analysis, we will make use of the result on the minimum singular value of a Gaussian perturbed square matrix~\citep{sankar2006smoothed}. However, since the submatrix of the augmented constraint matrix $\augconstraints_b$ could contain rows from the identity matrix $I_m$ or the all-one vector $\vecone_m$ (which are not perturbed by Gaussian noise), we need to perform some special treatments to these cases. We prove \Cref{thm:smoothed-analysis} in \Cref{app:smooth}.

\newpage
\bibliographystyle{ACM-Reference-Format}
\bibliography{ref}

\newpage
\appendix

\startcontents[appendix]
\clearpage
  \begin{center}\bfseries Appendix Table of Contents\end{center}
  \printcontents[appendix]{l}{1}{\setcounter{tocdepth}{2}}

\section{Comparing local Stackelberg benchmark with other benchmarks}
\label{app:benchmarks}

We will compare the local Stackelberg with other solution concepts in this section. We will compare solution concepts according to two criteria.

The first is the utility the principal achieves in a solution concept. Since there can be multiple solutions in a solution concept, we will compare the least principal utilities across solution concepts. 

The second criterion is the computational feasibility of approximating that solution concept through interactions with a learning agent. 

\begin{figure}[htbp]
\centering
\begin{tikzpicture}[scale=0.5,very thick,draw=blue,
every node/.style={text=black}]
  \coordinate (A) at (0,0);      %
  \coordinate (B) at (6,0);      %
  \coordinate (C) at (3,6);      %
  \coordinate (D) at (3,0);      %

  \coordinate (F) at (2,0);      %
  \coordinate (G) at (4,0);      %
  \coordinate (E) at (3,3);      %

  \draw (A) -- (B) -- (C) -- cycle;   %
  \draw (C) -- (E);                   %
  \draw (F) -- (E) -- (G);            %

  \node at (3,1)   {$P_1$};
  \node at (1.6,2) {$P_2$};
  \node at (4.4,2) {$P_3$};

\fill (G) circle[radius=5pt];   %
\fill (E) circle[radius=5pt];   %

\node[below right]  at (G) {$x^*$};   %
\node[above right] at (E) { $\overline{x}$};
\end{tikzpicture}
\caption{LSE vs Smoothed LSE : A game where $\overline{x}$ is a smoothed LSE for small enough $\eta$ and $x^*$ is the sole LSE. $\overline{x}$ is the optimal strategy within polytopes $P_2$ and $P_3$. And $x^*$ is the optimal strategy within $P_1$. The strategy $\overline{x}$ with action of $P_1$ has utility 1 more than the strategy of $\overline{x}$ with actions of $P_2$ or $P_3$. Since $P_1$ is a very thin polytope, $\overline{x}$ is nevertheless a smoothed LSE. It is however not an LSE.}
  \label{fig:thin-polytope}
\end{figure}
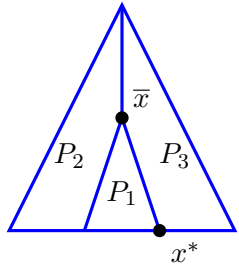

\begin{itemize}[leftmargin=1em]
    \item \textbf{Stackelberg equilibrium:} The Stackelberg equilibrium is the local Stackelberg equilibrium with the highest utility for the principal.  However, this can be intractable to approximate when interacting with a mean-based learner~\citep{brown2024learning}.  
    \item \textbf{Coarse Correlated Equilibrium (CCE):} A CCE be efficiently approximated against a mean-based learner. This can be done by simply employing a no-regret learning algorithm. There are cases where the minimum principal utility among CCEs is greater than the minimum principal utility among local Stackelberg equilibria and cases where the relation goes the other way. In particular, for Stackelberg security games, the local Stackelberg equilibrium achieve the same utility for the principal as the global Stackelberg equilibrium (see \Cref{prop:SSG} for more details). Hence, for this class of games, the local Stackelberg equilibrium yields a higher utility compared to any CCE~\citep{von2010leadership}. 
    \item \textbf{Smoothed Local Stackelberg Equilibrium.} The smoothed local Stackelberg can be thought of the local Stackelberg equilibrium in the setting where there is some noise (from $\mathcal{N}(0, \eta^2I)$) added to the principal's strategy.

    The noise added allows us to side-step the challenge of discontinuous utilities for the principal. However, the utilities of the smoothed local Stackelberg can be much worse than the utility of any local Stackelberg.

    The smoothed local Stackelberg is defined as follows:
    
    \begin{definition}[($\varepsilon, \delta, \eta$)-Smoothed Local Stackelberg Equilibria]
    A principal's strategy $x \in \Delta(\cA)$ is an ($\varepsilon, \delta, \eta$)-Smoothed Local Stackelberg strategy if \[
   \forall \x'\in\ball_1(\x;\delta),\quad 
   \E_{z' \sim \mathcal{N}(\x', \eta^2I)}\sup_{y'\in\BR(\z')}\utilp(\z', y') \leq \E_{z \sim \mathcal{N}(\x, \eta^2I)} \sup_{y\in\BR(\z)}\utilp(\z, y)+\eps\delta,
    \]
    where $\ball_1(\x,\delta)$ denote the $\ell_1$ ball of radius $\delta$ around $\x$, i.e., the set of strategies with $\|\x'-\x\|_1\le\delta$.
        
    \end{definition}

    The random noise added to the principal's strategy smoothens the principal's utility and makes it continuous. As a result, gradient-free optimization methods~\citep{maheshwari2023convergent, fiez2020implicit, Flaxman2004OnlineCO, nesterov2017random, gasnikov2023randomized} can be used to compute a ($\varepsilon, \delta, \eta$)-Smoothed Local Stackelberg equilibrium.

    We can however show that the worst smoothed local Stackelberg equilibrium can have principal's utility at least $\Omega(\varepsilon^{1/d})$ worse than the the worst local Stackelberg equilibrium.

    We can show this by constructing an example as in \Cref{fig:thin-polytope}. In this example, there is a very thin polytope $P_1$ that still has a ball of radius $R$. Note that this example satisfies the assumptions we use for our results.
    
    At the point of intersection of the three polytopes $\overline{x}$, the actions of polytopes $P_1$, $P_2$, $P_3$ yield the principal utilities in that order, with the utilities differing by a constant amount. $\overline{x}$ is the optimal strategy within $P_2$. Since the action of $P_3$ yields lesser utility and since $P_1$ is a very thin polytope, $\overline{x}$ is a smoothed LSE when $\eta \in o(R\varepsilon^{1/d})$.

    $\overline{x}$ is however not an LSE. $\x^*$ is the only LSE and has at least $\Omega(R)$ utility larger than $\overline{x}$.

    On the other hand, if $\eta \in \Omega(\varepsilon^{1/d})$, two points in the same best-response polytope at $\ell_1$ distance $\eta$ both form a smoothed Stackelberg equilibrium. By linearity of utilities within best-response polytopes, one of these points has principal utility less than $\Omega(\eta)$ compared to the other.

    \nacomment{Say that this is tractable. But to achieve the same utility as local stackelberg, we need exponentially smaller approximation }
\end{itemize}

\paragraph{Stackelberg Security Games.}
In the remainder of this section, we consider a structured class of games: the Stackelberg Security Games (SSG)~\citep{kiekintveld2009computing,an2012security}. We will show that under standard non-degeneracy assumptions (\Cref{assump:SSG}), the value of any local Stackelberg equilibria is the same as the value of the global (strong) Stackelberg equilibria.

We set up the notations for the SSG model, following the formulation from \citep{haghtalab2022learning}.
In a SSG, there is a set of $n$ targets that the principal aims to protect. The agent has action space $\aspacea=[n]$, i.e., the agent can choose (potentially randomly) a target to attack. The principal's strategy space is a downward-closed subset $\aspacep\subseteq [0,1]^n$, where each coordinate $x_i$ represents the amount of resource that the principal puts in protecting target $i\in[n]$. For $\x\in\aspacep$ and $y\in\aspacea$, the principal's and agent's utility functions are $\utilp(\x,y)=u^y(x)$ and $\utila(\x,y)=v^y(x)$, where $u^y$ and $v^y$ are 1-dimensional functions that are strictly increasing and strictly decreasing respectively, and satisfy the slope bounds in \Cref{assump:SSG}.

\begin{assumption}[Regularity~\citep{haghtalab2022learning}]
\label{assump:SSG}
    There exists a constant $C\ge1$ such that for all $0\le s<t\le1$ and $y\in\aspacea$, the functions $u^y$ and $v^y$ satisfy
    \begin{align*}
        \frac{1}{C}\le\frac{v^y(s)-v^y(t)}{t-s}\le C,
        \quad
        0 < \frac{u^y(t)-u^y(s)}{t-s}\le C.
    \end{align*}
\end{assumption}

\begin{proposition}[Equivalence of Local and Global Stackelberg Values]
\label{prop:SSG}
    For Stackelberg Security Games satisfying the regularity assumptions in \Cref{assump:SSG}, every local Stackelberg equilibria $\tx$ achieves the same principal utility as the global Stackelberg equilibria $\xst$, i.e.,
    \begin{align*}
        \utilp(\tx,\BR(\tx))=\utilp(\xst,\BR(\xst)).
    \end{align*}
\end{proposition}

\begin{proof}[Proof of \Cref{prop:SSG}]
    We prove this proposition using the characterization through \emph{conservative strategies} proposed by \citet{haghtalab2022learning}. A principal's strategy $\x\in\aspacep$ is called \emph{conservative} if for all $y\in[n]$, $x_y>0$ only if $y\in\BR(\x)$, i.e., $\x$ only protect targets that belongs to the agent's best response set. Following \citep[Proposition 3.5]{haghtalab2022learning}, it is without loss of generality to assume that both $\tx$ and $\xst$ are conservative strategies, as otherwise they can be easily transformed to conservative strategies without changing the principal's utility under best response.

    Since both $\xst$ and $\tx$ are conservative principal strategies, Proposition 3.5 from \citep{haghtalab2022learning} show that 
    $\xst$ is the unique conservative maximizer of $\utilp(\xst,\BR(\xst))$, which is also the unique conservative minimizer of $\utila(\xst,\BR(\xst))$. Therefore, if 
    if $\utilp(\tx,\BR(\tx))< \utilp(\xst,\BR(\xst))$, it must be the case that $\utila(\tx,\BR(\tx))>\utila(\xst,\BR(\xst))$. Therefore, from Lemma 3.4 of \citep{haghtalab2022learning}, we have $\BR(\tx)\subseteq \BR(\xst)$, and $\tx_y<\xst_y$ for all targets $y\in\BR(\xst)$.
    Since the principal's action space $\aspacep$ is downward-closed, 
    we can therefore infinitesimally increase the resource that the principal invests on every $y\in\BR(\tx)$ without changing the best response set. In other words,
    there exists a vector $\Vec{\eps}\le \xst-\tx$ such that $\eps_y>0$ for all $y\in\BR(\tx)$, such that the new strategy $\tx'$ satisfies $\BR(\tx')=\BR(\tx)$. Since $u^y$ is strictly increasing, we have
    \begin{align*}
        \utilp(\tx',\BR(\tx'))=\max_{y\in\BR(\tx')} u^y(\tx_y')
        =\max_{y\in\BR(\tx)} u^y(\tx_y')
        >\max_{y\in\BR(\tx)} u^y(\tx_y)
        =\utilp(\tx,\BR(\tx)),
    \end{align*}
    which contradicts with the fact that $\tx$ is a local Stackelberg equilibria! Therefore, any local Stackelberg equilibria must induce the same principal utility as the global Stackelberg equilibria. 
\end{proof}

\section{More details on the assumptions}
\label{app:assumptions}
In this section, we provide more discussions on our assumptions in \Cref{sec:assumptions} and their implications.

\begin{remark}[Remarks on \Cref{assump:far-from-boundary}]\label{remark:far-from-boundary}
    \Cref{assump:far-from-boundary} is an arguably weaker condition than the assumption made in previous works on computing an approximate Stackalberg Equilibrium from Best-Response oracle or from interactions with an agent, such as~\cite{blum2014learning,letchford2009learning,
    haghtalab2022learning}. Their assumption states that any polytope $\polytope_b$ contains an $\ell_2$ ball whose radius is lower bounded by some constant value $r_0$. \Cref{assump:far-from-boundary} is weaker for the following reasons:
    \begin{enumerate}
        \item 
        \Cref{assump:far-from-boundary} is a direct implication of the ball assumption.
        If a polytope $\polytope_b$ contains a point $\x_b$ that satisfies the ball assumption, then the distance from $\x_b$ to all boundaries of $\polytope_b$ is at least $r_0$, which naturally implies that the minimum coordinate of $\x_b$ is also lower bounded by $r_0$.
        \item \Cref{assump:far-from-boundary} only concerns the distance to the simplex boundaries, whereas the ball assumption requires the distance to all polytope boundaries to be lower bounded as well.
        \item Previous works that build on the ball assumption (e.g., \cite{blum2014learning,letchford2009learning,haghtalab2022learning,haghtalab2024calibrated}) usually incurs a dependency on the volume of the ball, which can be exponential in the dimension (i.e. the number of principal's actions). In contrast, our bound only has a polynomial dependence on $1/\rinf$.
    \end{enumerate}
\end{remark}

We then discuss how \Cref{inf_assumption:polytopes-far} can be implied by a lower bound on the singular values of constraint matrices, and justify this lower bound in the smoothed analysis framework.

\begin{definition}[Constraint Matrix]
    \label[definition]{def:constraint-matrix}
    For each of agent's action $b\in\aspacea$, let $\constraints_b$ be the \emph{constraint matrix} formed by the set of potential hyperplanes that separate $\polytope_b$ from all other polytopes---i.e., $\constraints_b=\begin{bmatrix}
        \h_{b,b'}^{\transpose}
    \end{bmatrix}_{b'\in\aspacea\setminus\{b\}}$.
    Additionally, 
    since the strategies all satisfy the simplex constraint (i.e., $\vecone^{\transpose}\x=1$ and $\x\ge\veczero$), we
    define the \emph{augmented constraint matrix} as the matrix obtained by augmenting $\constraints_b$ with an all-$1$ row vector ($\vecone_m$) and the identity matrix ($I_m$), and denote it with
    $\augconstraints_b=\begin{bmatrix}
        \constraints_b^{\transpose}&
        \vecone_m^{\transpose}&
        I_m
    \end{bmatrix}^{\transpose}$. 
\end{definition}

\begin{assumption}[Minimum singular value of square submatrix of $\augconstraints_b$]
    \label[assumption]{assump:singular-value}
    Let $\sqsubmatrix_m(\augconstraints_b)$ denote all the $m\times m$ square submatrices of $\augconstraints_b$.
    We assume that for all $b\in\aspacea$ and all square submatrices $\augconstraints'_b\in\sqsubmatrix_m(\augconstraints_b)$, the minimum singular value of $\augconstraints'_b$ satisfies $\sigma_{\min}(\augconstraints_b')\ge\sigmalb$.
\end{assumption}
We will justify Assumption~\ref{assump:singular-value} in \Cref{sec:smoothed-analysis}, by showing that it holds with high probability under the smoothed analysis framework of~\citet{spielman2004smoothed}.

\paragraph{Structural properties implied by \Cref{assump:singular-value}}

In the following lemma, we show that \Cref{assump:singular-value} implies that the polytopes and hyperplanes are sufficiently separated from each other, which is a formal version of \Cref{inf_assumption:polytopes-far}. We defer the proof of this lemma to \Cref{app:implication}.

\begin{lemma}[Polytopes/Hyperplanes are Far Apart]
    \label[lemma]{lemma:polytopes-far}
    Let $\sigmalb$ be the lower bound on the minimum singular values defined in \Cref{assump:singular-value}. 
    We have
    \begin{enumerate}
        \item Let $R_1=\sigmalb/(2m^{3/2})$. For all $\x\in\Delta(\aspacep)$, $|\surroundingP_{R_1}(\x)|\le m$, i.e., there are at most $m$ polytopes that have distance at most $R_1$ to $\x$.
        \item 
        Let $R_2=\sigmalb/(2m)$.
        For all $b\in\aspacea$ and all $\x\in\polytope_b$, there are at most $(m-1)$ rows $h$ of $\augconstraints_b$ that satisfy $0 \le \langle x,h\rangle \le R_2$. This implies that there are at most $m-1$ hyperplanes from that polytope that $\x$ can be very close to.
    \end{enumerate}
    
\end{lemma}

\nacomment{Add assumption about lower bound on Lipschitzness of agent utilities that we need for mean-based learners.}

\section{Querying through Average Strategies
}\label{sec:query_average_strategies}
Recall that we use $\x^{(t)}$ to denote the principal's strategy at round $t$, and $\avgx^{(t)}=\frac{1}{t}\sum_{s=1}^t \x^{(s)}$ to denote the average strategy during the first $t$ rounds. Since a fictitious play agent best responds to the average strategy, we treat $\avgx^{(t)}$ as our \emph{effective search point} at round $t$. Using the $\broracle$, we can find $\BR(\avgx^{(t)})$ as long as $\avgx^{(t)}$ is far enough from polytope boundaries, which our algorithm design enforces. Hence we can infer which best response polytope the average strategies lie in. This best response further dictates which utility function $\up(\cdot,y_t)$ the principal should be optimizing. 

However, an inherent constraint of the \emph{effective search points} is that we can not move too far between consecutive search points. Specifically, the maximum distance that we can travel in one step (i.e., $\|\avgx^{(t+1)}-\avgx^{(t)}\|$) shrinks at a rate of $O(1/t)$, and is often constrained by the previous search point's distance to the simplex boundaries. To capture this, we say that the principal takes \emph{step size} $\eta^{(t)}$ at round $t$ if the average strategy moves an $\ell_1$ distance of $\eta^{(t)}$, i.e.,
\[
\|{\avgx}^{(t)} - {\avgx}^{(t-1)}\|_1 = {\eta^{(t)}}/{t}.
\tag{Step size $\eta^{(t)}$}
\]

In \Cref{alg:move-one-step}, we introduce a procedure for updating the average strategy with a pre-specified update direction and step size.

\begin{algorithm}[t]
  \caption{$\moveonestep$}
  \label{alg:move-one-step}
  \Input{Current round $t$; current average principal strategy $\bar{\mathbf{x}}^{(t-1)}\in\mathcal{X}$; 
         move direction $\mathbf{u}^{(t)}\in\mathcal{X}$; 
         step size $0\le \eta^{(t)}\le\|\mathbf{u}^{(t)}-\bar{\mathbf{x}}^{(t-1)}\|_1$.}
  \Output{Principal strategy $\mathbf{x}^{(t)}$ such that the average moves by $\eta^{(t)}$ in the direction of $\mathbf{u}^{(t)}$.}

  \[
  \mathbf{x}^{(t)} =
  \left(1-\frac{\eta^{(t)}}{\|\mathbf{u}^{(t)}-\bar{\mathbf{x}}^{(t-1)}\|_1}\right)\bar{\mathbf{x}}^{(t-1)}
  + \frac{\eta^{(t)}}{\|\mathbf{u}^{(t)}-\bar{\mathbf{x}}^{(t-1)}\|_1}\,\mathbf{u}^{(t)}.
  \]
\end{algorithm}

\Cref{alg:move-one-step} chooses a point $\x^{(t)}$ on the line segment between $\avgx^{(t-1)}$ and $\u^{(t)}$. In other words, $\x^{(t)}$ is formed by moving the average strategy $\avgx^{(t-1)}$ in the direction of $\u^{(t)}$. Note that $\x^{(t)}$ is a valid strategy in the simplex $\cX$ because it is a linear combination of two valid strategies $\avgx^{(t-1)}$ and $\u^{(t)}$. It has the following property:
\begin{align*}
     \avgx^{(t)} - \avgx^{(t-1)} &= \frac {\x^{(t)} - \avgx^{(t-1)}} t =
     \frac{\eta^{(t)}}{t} \cdot\frac{\u^{(t)} - \avgx^{(t-1)}}{\|\u^{(t)} - \avgx^{(t-1)}\|_1 }
     \quad\Rightarrow\quad
     \|{\avgx}^{(t)} - {\avgx}^{(t-1)}\|_1 = \frac{\eta^{(t)}}{t}.
\end{align*}

\begin{remark}[Maximum step size in any direction]
\label[remark]{remark:max-step-size}
    At round $t$, the maximum $\ell_1$ that the principal can move its average strategy towards the direction $\u^{(t)}$ is $\|\u^{(t)} - \avgx^{(t-1)}\|_1/t$. In other words, the maximum feasible step size is $\|\u^{(t)} - \avgx^{(t-1)}\|_1$.
\end{remark}
\usetikzlibrary{decorations.pathreplacing}

\begin{figure}[htbp]
    \centering
    \begin{tikzpicture}[scale=0.8, every node/.style={font=\footnotesize}]

    \coordinate (xtm1) at (0,0);
    \coordinate (xt) at (2,0);
    \coordinate (ut) at (6,0);
    \coordinate (xnext) at (5.3,0);
    
    \draw[thick] (xtm1) -- (ut) node[right] {};

    \fill (xtm1) circle (2pt);
    \fill (xt) circle (2pt);
    \fill (ut) circle (2pt);
    \fill (xnext) circle (2pt);

    \node[below] at (xtm1) {$\avgx^{(t-1)}$};
    \node[below] at (xt) {$\avgx^{(t)}$};
    \node[below] at (ut) {$\u^{(t)}$};
    \node[below] at (xnext) {$\x^{(t)}$};

    \draw[decorate,decoration={brace,amplitude=8pt,mirror},red] (0,-0.5) -- (2,-0.5) 
        node[midway,below,yshift=-8pt,align=center] 
        {$\ell_1$ distance 
        $=\frac{\eta^{(t)}}{t}\le \frac{d}{t}$ };

    \draw[decorate,decoration={brace,amplitude=8pt}] (0,0.2) -- (6,0.2) 
        node[midway,above,yshift=8pt] 
        {$d$};

    \fill[red,opacity=0.4] (0,0.15) rectangle (2,-0.15);

\end{tikzpicture}

    \caption{An illustration of how to move the average strategies in $\moveonestep$. If $\|\avgx^{(t-1)} - \u^{(t)}\|_1=d$, then by choosing appropriate $\x^{(t)}$ along the line segment between $\avgx^{(t-1)} $ and $ \u^{(t)}$, the principal's average strategy can move $\ell_1$ distance of $\|\avgx^{(t-1)} - \avgx^{(t)}\|_1\in[0,\frac{d}{t}]$ (the shaded region is achievable).}
    \label{fig:move-one-step}
\end{figure}
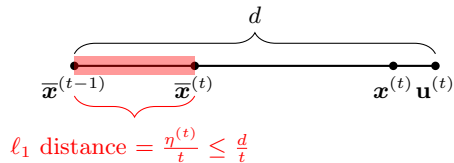

\subsection{Best-response oracle}\label{sec:br_oracle}

\paragraph{Best-response oracle.}
We implement a best-response oracle that when given $\avgx^{(t)}$ at distance at least $\alpha$ away from any polytope boundary, returns $\BR(\avgx^{(t)})$ with probability at least $1-1/t^2$.

If $\avgx^{(t)}$ is $\alpha$-far from boundaries, then according to the argument in \citep[Lemma F.3]{haghtalab2022learning}, the agent's average utility for action $\BR(\avgx^{(t)})$ is at least $\alpha\Delta$ higher than all other actions. Due to the mean-based condition, the agent's response $y^{(t)}$ is $\BR(\avgx^{(t)})$ with probability $\ge 1 - n \alpha \Delta$. 

The $\broracle$ chooses the strategy $\avgx^{(t)}$ for $2 \log t / (1 -  n \alpha \Delta)^2$ rounds after round $t$. And then it returns the most frequently occurring response. This response is the true best-response of $\avgx^{(t)}$ with probability $\ge 1 - 1/t^2$.

\paragraph{Detecting best-response change.} We detect crossing a boundary by moving from $\avgx$ to $\avgy$ if $\broracle(\avgx) \neq \broracle(\avgy)$. We have two possible cases: the best-responses are indeed different for $\avgx$ and $\avgy$. Or $\broracle$ incorrectly determines that the best-responses are different. This still means that one of $\avgx, \avgy$ is close to the boundary. So we already meet our goal of finding a point close to the boundary, which is the end-goal of detecting changes in best-responses.

\section{Supplementary materials for $\optwithinsubspace$}

\subsection{Correctness of $\optwithinsubspace$}
\label{sec:within-correctness}
In this section, we establish the correctness of $\optwithinsubspace$.
The $\optwithinsubspace$ subroutine invokes another subroutine $\searchpolytope$ that to find the hyperplane separating two points in different polytopes. 
We will defer the analysis of the $\searchpolytope$ subroutine to \Cref{sec:search-polytopes} and simply assume that given $\x, \x' \in \cX$ that lie in two different best response polytopes, the subroutine returns an $\alpha$-approximate hyperplane to the true separating hyperplane. That is, if the true hyperplane separating $\x,\x'$ is $\h$, then, $\searchpolytope$ returns a hyperplane $\hat{\h}$ that satisfies $\|\hat{\h} - \h\|_2 \leq \alpha$.

The main technical lemmas are listed below.

    \begin{restatable}[Closeness of optimal values within true and estimated polytopes]{lemma}{lemmaPPhatclose}
    \label[lemma]{lemma:estimated-polytope-close}
    Let $H, \hat{H} \in \mathbb{R}^{k \times m}$ with $k \leq n$ be the true and estimated constraints, which satisfy $\|H - \hat{H}\|_2 \leq \alpha \sqrt{m}$, where $\alpha$ is the estimation error that can be chosen sufficiently small.
    Consider polytopes $\polytope_b$ and $\hat{\polytope}_{b}$ defined as follows:
    \begin{align*}
        \polytope = \{\x \in \br^m: Hx \geq 0, \vecone^{\transpose}\x=1, \x\ge\veczero\}, \quad
        \hat{\polytope} = \{\x \in \br^m: \hat{H}x \geq \alpha, \vecone^{\transpose}\x=1, \x\ge\gamma\cdot\vecone \} 
    \end{align*}
    Then, when maximizing the principal's utility $\utilp(\cdot,b)$ over $\polytope_b$ and $\hatpolytope_b$, the corresponding optimal values satisfy
    \[\max_{\x \in \hat{P}} \up(\x, b) \leq \max_{\x \in P} \up(\x, b)  + \frac {2 \alpha {m}} {\sigmalb - \alpha \sqrt{m}}+2\gamma/\rinf,\]
    where $\alpha$ is chosen to be at most $O\left(\frac{\sigmalb}{m^2n}\right)$, $\sigmalb$ is the lower bound on minimum singular values from \Cref{assump:singular-value},
    $\rinf$ is the parameter from \Cref{assump:far-from-boundary},
    and $R_2=\sigmalb/m$ is from \Cref{lemma:polytopes-far}.
\end{restatable}

\begin{proof}[Proof sketch (Full proof in \Cref{app:proofPPhatsmall})]
    The high level idea is that given a point $\x$ satisfying constraints according to the constraint matrix $H\x \geq 0$, we want to perturb $\x$ to the point $\x+\z$ satisfying the slightly perturbed constraint of $\hat{H}\x \geq \alpha$.

    The constraints that $\x$ satisfies by a low margin are the most sensitive to being violated by perturbing $\x$. So we will focus on these constraints. Specifically, we will focus on the set of constraints indexed by $J$ that are satisfied by a margin of less than $\rtwo$. By our \Cref{lemma:polytopes-far}, we know that $|J| \leq m$. 

    Finding a $\z$ that satisfies the perturbed constraint with $\leq m$ constraints perturbed amounts to solving a system of equations with fewer equations than constraints. The norm of the minimum norm solution is provided by the inverse of the minimum singular value of $H_J$ which we assume to be bounded below by $\sigmalb$ (\Cref{assump:singular-value}).

    We can show that we can find a $\z$ with norm small enough that the remaining contraints outside of $J$ that were satisfied by a large margin remain satisfied even after perturbation by $\z$. 
\end{proof}

\subsection{Sample complexity of $\optwithinsubspace$}
\label{sec:within-complexity}

Recall that the time steps that the principal spends in $\optwithinsubspace$ can be divided into two categories: the \emph{improvement steps} and the \emph{searching steps}. In this section, we establish guarantees for the two categories separately. 

\paragraph{Analysis of the improvement steps.}
We will first analyze the \emph{improvement steps} by showing that 
each step improves the principal's utility by at least $\eps\delta/t$ (\Cref{lemma:improvement-steps}). This lemma will be crucial for bounding the total number of different polytopes, as the cumulative utility improvement is at most $1$.

\begin{lemma}
    [Utility Improvement]
    \label[lemma]{lemma:improvement-steps}
    Suppose $\avgx^{(t-1)}\to\avgx^{(t)}$ is an improvement step, where both strategies are inside $\polytope_b$. The principal's utility increases by at least $\frac{\eps \delta}{t}$ during this improvement step.
    \begin{align*}
        \utilp(\avgx^{(t)},b)
        -\util(\avgx^{(t-1)},b)
        \ge \frac{\eps\delta}{t}.
    \end{align*}
    
\end{lemma} 
\begin{proof}[Proof of \Cref{lemma:improvement-steps}]
    Since the $\optwithinsubspace$ algorithm does not terminate at round $t-1$, we know that there exists $\z\in\hatpolytope_{b}$ such that
    $
    \utilp(\z,b)>
    \utilp(\avgx^{(t-1)},b)+\eps\delta.
    $
    Therefore, by choosing $\x^{(t)}=\z$, the principal's utility at $\avgx^{(t)}$ can be calculated as
    \[
    \utilp(\avgx^{(t)},b)=
    \frac{t-1}{t}\cdot \utilp(\avgx^{(t-1)},b)
    +\frac{1}{t}\cdot\utilp(\z,b).
    \]
    The improvement in utility therefore satisfies
    \[
    \utilp(\avgx^{(t)},b)
        -\util(\avgx^{(t-1)},b)
    =\frac{\utilp(\z,b)-
    \utilp(\avgx^{(t-1)},b)}{t}
    >\frac{\eps\delta}{t}.
    \]

    This increase in utility happens over $O(\log t)$ rounds since the best-reponse oracle (\Cref{sec:br_oracle}) repeats strategies $O(\log t)$ times to find the best-response with high probability. Hence the improvement per round is $\ge \varepsilon \delta / (t \log t)$.
\end{proof}

\paragraph{Analysis of the searching steps.}

The searching phases consists of two subroutines: $\binarysearch$ (\cref{alg:binary-search}) and $\searchpolytope$ (\Cref{alg:search-for-polytope}, which we analyze in \Cref{sec:search-polytopes-correctness}).
We provide the guarantee of $\binarysearch$ and defer its proof to \Cref{app:binarysearch}.

\begin{lemma}[Binary search to get close to the boundary]
\label[lemma]{lem:BinarySearchUpperBound}
Let $\gamma$ be the minimum distance from all search point to the boundary in \Cref{lemma:estimated-polytope-close}.
Suppose the principal's average strategy moves from  $\avgx^{(t-1)}$ in polytope $\polytope_b$ to $\avgx^{(t)}$ in a different polytope $\polytope_{b'}$, with $\ell_1$ step size of $\|\avgx^{(t)}-\avgx^{(t-1)}\|_1=\frac{\eps}{t}$, where we assume $t\ge10\frac{\eps}{\gamma}$.
Then the procedure $\binarysearch$ (see \Cref{alg:binary-search}) takes at most
$s \leq O\Bigl(\frac{\varepsilon}{\gamma}+\log \frac {\varepsilon} {\alpha t}\Bigr)$
steps and returns an average strategy $\avgx^{(t+s)}$ that is within $\alpha$ $\ell_1$ distance close to the boundary.
\end{lemma}

\section{Supplementary materials for $\searchpolytope$}
\subsection{Correctness of $\searchpolytope$}

\label{sec:search-polytopes-correctness}

\thmsearchcorrectness*

In this section, we outline the main components of the proof of \Cref{thm:search_correctness}. We will present the full proof in \Cref{proof:search_correctness}.
    
Let $J$ of size $j$ index the polytopes already discovered in previous iterations. We will analyze the iteration of the algorithm with the search space $\hatsubspace_J = \{\x \in \sspacep: \hat{\h}_J \x \geq \alpha_j \}$, where $\hat{\h}_J$ is the matrix of approximations to the hyperplanes discovered thus far. Inductively we will show that if $\|\hat{\h}_{b,b_j} - {\h}_{b,b_j}\| \leq \alpha_j$ for every $j \in J$, the next constructed hyperplane satisfies $\|\hat{\h}_{b,\bnew} - h_{b, \bnew}\| \leq \alpha_j O(\alpha_j^2 m^2 / \sigmalb^2)$. 
By setting $\alpha_1 \leq \alpha (\sigmalb / m)^m$, we ensure that all $\alpha_j \leq \alpha$. This is because the number of iterations of this algorithm is at most $m$ as shown in the following lemma.

\begin{restatable}[Algorithm terminates after $m-1$ hyperplanes added]{lemma}{lemmaterminate}
\label{lem:search_poly_max_poly}
The $\searchpolytope$ algorithm will terminate after at most $m-1$ hyperplanes are added.
\end{restatable}

Now we will discuss how random search on the search space via Gaussian random vectors will discover new polytopes in $\mathcal{P}_{\rho}(\x^*)$ that are not yet discovered through sufficiently many search points landing in them.

\begin{lemma}[Search finds points close to a new boundary]
    With high probability, $\Omega(1/m)$ of the search points generated lie in a new, undiscovered polytope $\polytope_{b_{\mathrm{new}}}$ and lie at distance $\Theta(\eta_j \sqrt{m} + \alpha_j m^3/\sigmalb)$ of $\h_{b, b_{\mathrm{new}}}$. 
\end{lemma}
\begin{proof}
First we will show that many search points fall in a new, previously undiscovered polytope. In the space $\{\x \in \sspacep: {H_J} \leq \alpha_j\}$, the agent gets the same utility by playing any action in $J$ or the action of the current polytope. Since our search space is an approximation of this space, the agent is approximately indifferent among all already discovered actions for strategies in the search space (\Cref{lem:appx_indifference}). 

As long as there is a polytope yet to be discovered in $\mathcal{P}_{\rho}$, we show that random search generates a constant fraction of principal strategies where the action corresponding to undiscovered polytope yields a higher utility, indicating that the searched strategies lie in a previously undiscovered polytope (\Cref{lemma:discover_new_poly}).  

Finally, due to the $\eta$ scaling of the search vectors, the searched strategies are $\eta \sqrt{m}$ close to the point $\hat{\x}_J$ we search from which in turn is close to $\x^*$ (\Cref{lem:search_point_close}) which lies on $\h_{b, \bnew}$. This shows that the points discovered in $\polytope_{\bnew}$ are close to the boundary hyperplane $\h_{b, \bnew}$. Additionally, they are not too close and so we can recover their best-response through $\broracle$(\Cref{sec:br_oracle}). \end{proof}

The properties of the search vectors in the new polytope --- a significant fraction of all search vectors and lying close to the boundary allow us to accurately reconstruct $\h_{b, \bnew}$. Let $Y$ be a matrix of the search points that land in the new polytope $\polytope_{\bnew}$. We will construct a hyperplane solving $\hat{\h} = \argmin_{\h} \|\h Y^\transpose\|$. We will now show that $\hat{\h}$ is a good approximation to $\h_{b, b_{\mathrm{new}}}$

\begin{lemma}[Approximating a hyperplane]\label{lem:separating_hyperplane}
Suppose $\mathsf{RandomSearch}$ in the  search space $\hatsubspace_J = \{\x : \hat{\h}_J\x = \alpha_j\}$ generates $d \in \Theta(m^2 \log m)$ points.
Let $\bnew$ be the undiscovered polytope with the maximum number of search points. Let $Y \in \mathbb{R}^{k \times m}$ be a matrix where the rows are all search points in $\polytope_{\bnew}$ with distance from $\hat{\x}_J \in O(\eta \sqrt{m})$.

      Then $\hat{\h} = \argmin_{\h} \|hY^t\|_2$ satisfies $\|\hat{\h} - h_{b, b_{\mathrm{new}}}\| \leq \alpha$.  
\end{lemma}
\begin{proof}
    The proof idea is to first provide a lower bound in the minimum singular value of $Y$. Because the rows in $Y$ lie close to $\h_{b, \bnew}$, $\|h_{b, \bnew} Y^t\|$ is small. Since $\hat{\h}$ minimizes the norm $\argmin_h \|\hat{\h}Y^\transpose\|$, $\|h_{b, \bnew} Y^t\|$ is also small. Due to the lower bound on the minimum singular value of $Y$, this will imply that $\|\hat{\h} - h_{b, \bnew}\|$ is small.
\end{proof}

So far we argued that as long as some polytope in $\mathcal{P}_{\rho}(\x^*)$ is undiscovered, we will find new polytopes through random search. To complete the argument, we will argue that this means that \emph{all} polytopes in $\mathcal{P}_{\rho}(\x^*)$ are discovered before the algorithm terminates. This is due to the property that no more than $m$ polytopes surround a point and the only way the algorithm could terminate without finding some polytope in $\mathcal{P}_{\rho}(\x^*)$ is if there were more than $m$ surrounding polytopes.

\subsection{Complexity of $\searchpolytope$}
\label{sec:search-polytopes-complexity}

\begin{lemma}[Number of rounds spent searching for polytopes]\label{lem:search_upper_bound}
    When $\alpha \in O(\gamma \sigmalb^2 / (n^2 m^7 \log m))$, the number of rounds spent in calls of $\searchpolytope$ is at most $O(n^2 m^2 \log m)$.
\end{lemma}

\begin{proof}
    $\searchpolytope$ is invoked at most $n^2$ times, once when encountering each hyperplane. The different types of movements are: 1) taking a random step of $\ell_2$ distance at most $\eta \sqrt{m}$, 2) returning back to the search point, and 3) projecting the search point back to updated search spaces.

    Each of these movements have a $\ell_1$ length in $\Delta \in O(\alpha m^5 / \sigmalb^2)$. And the total number of these movements is $r = O(n^2 m^2 \log m)$ over all the $n^2$ possible times $\searchpolytope$ is invoked.

    In rounds $t$ with $\gamma / t \geq \Delta$, we can make the movement in a single round. So the number of rounds before $t_0 = \gamma / \Delta$ is at most $r = O(n^2 m^2 \log m)$.

    For rounds after $t_0$, may need multiple steps to make the movement. In these rounds $t$, we will traverse exactly $\gamma / t$ $\ell_1$ distance.

    The total distance traversed while making all these movements is at most $r \Delta$. Let us lower bound the total distance traversed by the total distance traveled in rounds after $t_0 = \gamma / \Delta$.
    \begin{align*}
        r \Delta &\ge \sum_{\tau = t_0}^{t_0 + K} \frac \gamma \tau \ge \gamma \log \left (1 + \frac K {t_0} \right ) \quad
        \Longrightarrow K \le t_0 \left (\exp \left (\frac {r \Delta} \gamma  \right ) - 1 \right ) 
    \end{align*}
    Our assumption that $\alpha \in O(\gamma \sigmalb^2 / (n^2 m^7 \log m))$ implies that $r \Delta / \gamma \le 1.$ So, using $e^x - 1 \leq x$, $K \le t_0 \frac {r \Delta} {\gamma}$.
    Therefore the total number of rounds spent in $\searchpolytope$ is at most the number of rounds before $t_0$ which is $\leq r \in O(n^2 m^2 \log m)$ and the number of rounds after $t_0$ which is also $ \leq r \in O(n^2 m^2 \log m)$. 
\end{proof}

\section{Full proofs of results}

\subsection{Proof of \cref{lemma:polytopes-far}}
\label{app:implication}
\begin{proof}[Proof of \Cref{lemma:polytopes-far}, part (1)]
    For the sake of contradiction, assume that there are exists $\x\in\Delta(\aspacep)$ such that $\surroundingP_{R_1}(\x)$ contains at least $m+1$ polytopes. We denote them with $b_0,b_1,\ldots,b_m$, where we also assume without loss of generality that $\x\in\polytope_{b_0}$ (i.e., $b_0$ is a best response to strategy $\x$). 
    
    We will first show that for all $i\in[m]$, we have $0\le \h_{b_0,b_i}^{\transpose}\x\le \sqrt{m}$.
    Since $\distance(\x,\polytope_{b_i})\le $, there exists $\z_i\in\polytope_{b_i}$ such that $\|\x-\z_i\|_2\le $. This implies \[\utila(\x,b_0)-\utila(\z_i,b_0)=\ip{\util_{b_0}}{\x-\z_i}\le \|\util_{b_0}\|_2\cdot\|\x-\z_i\|_2\le \sqrt{m}.\]
    Therefore, we can bound the inner product of $\x$ and the hyperplane $\h_{b_0,b_i}=\util_{b_0}-\util_{b_i}$ as
    \[
     \ip{\h_{b_0,b_i}}{\x}=\utila(\x,b_0)-\utila(\x,b_i)
     =\underbrace{\utila(\x,b_0)-\utila(\z_i,b_0)}_{\le \sqrt{m}}+\underbrace{\utila(\z_i,b_0)-\utila(\z_i,b_i)}_{\le0 \text{ since }b_i\in\BR(\z_i)}
     \le \sqrt{m}.
    \]
    We also know $\ip{\h_{b_0,b_i}}{\x}\ge0$ since $b_0\in\BR(\x)$. As a result, we have $|{\h_{b_0,b_i}}^{\transpose}{\x}|\le \sqrt{m}$.

    Finally, we establish a contradiction using the following matrix $K\in\br^{m\times m}$, which is a square sub-matrix of the augmented constraint matrix $\augconstraints_{b_0}$.
    \[
    K=\begin{bmatrix}
        \h_{b_0,b_i} 
    \end{bmatrix}_{i\in[m]}.
    \]
    Combing the above bound on $|{\h_{b_0,b_i}}^{\transpose}{\x}|$ and the fact that $\|x\|_2\ge\frac{\|\x\|_1}{\sqrt{m}}=1/\sqrt{m}$, we have
    \[
    \frac{\|K\x\|_2}{\|\x\|_2}\le\sqrt{m}\cdot\sqrt{\sum_{i=1}^m  |{\h_{b_0,b_i}}^{\transpose}{\x}|^2}\le\sqrt{m}\cdot
    \sqrt{\sum_{i=1}^m  (\sqrt{m})^2}= m^{2/3}=\frac{\sigmalb}{2},
    \]
    which contradicts with the assumption that \[\sigma_{\min}(K)=\min_{\x\neq0}\frac{\|K\x\|_2}{\|\x\|_2}\ge\sigmalb.\]
    Therefore, we conclude that there cannot be more than $m$ polytopes in $\surroundingP_{}(\x)$ where $=\frac{\sigmalb}{2m^{2/3}}$.
\end{proof}

\begin{proof}[Proof of \Cref{lemma:polytopes-far}, part (2)]
    The second claim follows from a very similar proof. Suppose for the sake of contradiction that there exists a submatrix $K\in\sqsubmatrix_m(\augconstraints_b)$ of $m$ conditions, and an  $\x\in\polytope_b\subseteq\Delta(\aspacep)$, such that $\x$ satisfies all constraints in $K$ with margin at most $R_1$, then we have
    \begin{align*}
        \|K\x\|^2\le\sqrt{m\cdot R_1^2}=\sqrt{m}R_1\le mR_1\|\x\|_2,
    \end{align*}
    which contradicts with $\sigma_{\min}(K)\ge\sigmalb$ for $R_2=\frac{\sigmalb}{2m}$. Therefore, there can be at most $m-1$ constraints from $\augconstraints_b$ that are satisfied by a margin of at most $R_2$.
\end{proof}

\subsection{Proof of \Cref{lemma:estimated-polytope-close}}
\label{app:proofPPhatsmall}

\begin{restatable}{lemma}    
{lemma:estimated-polytope-close}
    Let $H, \hat{H} \in \mathbb{R}^{k \times m}$ with $k \leq n$ be the true and estimated constraints, which satisfy $\|H - \hat{H}\|_2 \leq \alpha \sqrt{m}$, where $\alpha$ is the estimation error that can be chosen sufficiently small.
    Consider polytopes $\polytope_b$ and $\hat{\polytope}_{b}$ defined as follows:
    \begin{align*}
        \polytope = \{\x \in \br^m: Hx \geq 0, \vecone^{\transpose}\x=1, \x\ge\veczero\}, \quad
        \hat{\polytope} = \{\x \in \br^m: \hat{H}x \geq \alpha, \vecone^{\transpose}\x=1, \x\ge\gamma\cdot\vecone \} 
    \end{align*}
    Then, when maximizing the principal's utility $\utilp(\cdot,b)$ over $\polytope_b$ and $\hatpolytope_b$, the corresponding optimal values satisfy
    \[\max_{\x \in \hat{P}} \up(\x, b) \leq \max_{\x \in P} \up(\x, b)  + \frac {2 \alpha {m}} {\sigmalb - \alpha \sqrt{m}}+2\gamma/\rinf,\]
    where $\alpha$ is chosen to be at most $O\left(\frac{\sigmalb}{m^2n}\right)$, $\sigmalb$ is the lower bound on minimum singular values from \Cref{assump:singular-value},
    $\rinf$ is the parameter from \Cref{assump:far-from-boundary},
    and $R_2=\sigmalb/m$ is from \Cref{lemma:polytopes-far}.
\end{restatable}

We prove this lemma by showing that the $\ell_1$ Hausdorff distance between $\polytope$ and $\hatpolytope$ are close, i.e.,
for any $\x \in \polytope$, we will construct a $\z\in\br^{m}$ such that $\x+\z \in \hat{\polytope}$ and $\|\z\|_1$ is small.

This will automatically imply the statement of the Lemma since the principal's utilities are $1$-Lipschitz in $\ell_1$ distance within the polytope. The utility within a polytope due to a strategy $\x$ is $\ip{\u_b}{\x}$. The absolute difference in utilities between two strategies is \[|\up(\x, b) - \ua(\y,b)| = |\ip{\u_b}{\x} - \ip{\u_b}{\y}| \leq \|\u_b\|_\infty \|\x-\y\|_1 \leq \|\x-\y\|_1\]

We bound the Hausdorff distance between $\polytope_b$ and $\hatpolytope_b$ by the sum of the Hausdorff distances between $\polytope_b$ and $\Tilde{\polytope}_b$ and between $\Tilde{\polytope}_b$ and $\hat{\polytope}_b$, where
\[\Tilde{\polytope} = \{\x \in \br^m: H\x \geq 0, \vecone^{\transpose}\x=1, \x\ge \gamma\cdot \vecone\}\]

Let us start by bounding the Hausdorff distance between $\polytope$ and $\Tilde{\polytope}$. 

\begin{lemma}
Assume $\gamma \le \rinf$ where $\rinf$ is the parameter from \Cref{assump:far-from-boundary}. The Hausdorff distance between $\polytope$ and $\Tilde{\polytope}$ is upper bounded by $2\gamma/\rinf$.
In other words, for every $\x \in \polytope$, there is a $\y \in \Tilde{\polytope}$ with $\|\x - \y\|_1 \leq 2\gamma / \rinf$. 
\end{lemma}
\begin{proof}
    By \Cref{assump:far-from-boundary}, there is a point $\x_0\in\polytope$ that satisfies $\x\ge \rinf\cdot\vecone$.
    Therefore, for any $\x \in \polytope$ and $\lambda \in [0,1]$, we construct $\y_{\lambda}$ as follows: $\y_{\lambda} = \lambda \x_0 + (1 - \lambda) \x.$
    Clearly, $\y_\lambda$ lies in $\polytope$ by the convexity of polytopes.
    Moreover, since $\x \ge 0$, we have $\y_{\lambda}\ge\lambda\x_0\ge(\lambda\rinf)\cdot\vecone$. We can thus choose $\lambda=\gamma/\rinf$ to ensure $\y_\lambda\in\Tilde{\polytope}$. Finally, for the distance between $\x$ and $\y_{\lambda}$, we have $\|\x - \y_{\lambda}\|_1 = \lambda\|\x-\x_0\|_1\le2\lambda=2\gamma/\rinf$.
\end{proof}

Next, let us bound the Hausdorff distance between $\Tilde{\polytope}$ and $\Hat{\polytope}$.

\begin{lemma}\label{lem:hausdorff}
When $\alpha\ll\frac{\sigmalb}{m^2n}$,
the Hausdorff distance between $\Tilde{\polytope}$ and $\Hat{\polytope}$ is upper bounded by $2\alpha\/\sigmalb$. In other words,
    for every $\x \in \Tilde{\polytope}$, there is a $\y \in \Hat{\polytope}$ with $\|\x - \y\|_1 \leq 4\alpha m/ \sigmalb$.
\end{lemma}

\begin{proof}
Given any $\x\in\Tilde{\polytope}$, we aim to construct $\z$ with small norm such that $\x+\z \in\hatpolytope$.
To make sure that $\x+\z\in\hatpolytope$,
we want $\z$ to satisfy the following two constraints:
(1) $\hat{H}(\x+\z) \geq 
    \alpha\cdot\vecone_k$; 
(2) $\vecone^{\transpose}(\x+\z) = 1$, i.e., $\vecone^{\transpose}\z = 0$.

 Let $J\subseteq[k]$ be the index of rows in $H$ that are satisfied by $\x$ by less than a $R_2$ 
 margin, where $\rtwo$ is the parameter from \Cref{lemma:polytopes-far}.
 Equivalently, $H_J$ contains all rows $\h$ such that $0\le \h^{\transpose}\x\le\rtwo$.
 \Cref{lemma:polytopes-far} guarantees that $|J| \leq m-1$. Let $H_{\setminus J}$ denote all other rows. That is, $\veczero \leq H_J \x \leq \rtwo\cdot\vecone, \text{ and } H_{\setminus J} \x \geq \rtwo\cdot\vecone$.

 We will first find a $z$ that satisfies constraint (2) and constraint (1), but only restricted to rows in $J$, i.e., $\hat{H}_J(\x+\z) \geq \alpha\cdot\vecone_{|J|}$.
 Defining $E_J = H_J - \hat{H}_J$ to be the error matrix of rows in $J$, and define $\hat{M}=\begin{pmatrix}
     \hat{H}_J \\ \vecone_m
 \end{pmatrix}$, we set $\z$ as follows:
 \[
 \z = \hat{M}^{\dagger}
         \begin{pmatrix}
         E_J \x + \alpha\vecone_{|J|} \\
         0
     \end{pmatrix}
     = \hat{M}^{\transpose}(\hat{M}\hat{M}^{\transpose})^{-1}
     \begin{pmatrix}
         E_J \x + \alpha\vecone_{|J|} \\
         0
     \end{pmatrix}.
 \]
 In the remainder of the proof, we will first show that $\z$ satisfies both constraints (1) and (2), then upper bound the norm $\|\z\|_2$.

 \paragraph{$\z$ satisfies constraints (2) and (1) restricted to $J$.}
    By our construction of $\z$, we have
    \begin{align*}
        \hat{M}\z=\hat{M}\hat{M}^{\dagger}
         \begin{pmatrix}
         E_J \x + \alpha\vecone_{|J|} \\
         0
     \end{pmatrix}=\begin{pmatrix}
         E_J \x + \alpha\vecone_{|J|} \\
         0
     \end{pmatrix}.
    \end{align*}
    Therefore, the vector $\x+\z$ satisfies
    \begin{align*}
        \hat{M}(\x+\z)=\begin{pmatrix}
         \hat{H}_J(\x+\z)\\
         \vecone^{\transpose}(\x+\z)
        \end{pmatrix}=\begin{pmatrix}
            \hat{H}_J\x\\
            \vecone^{\transpose}\x
        \end{pmatrix}+\begin{pmatrix}
         E_J \x + \alpha\vecone_{|J|} \\
         0
     \end{pmatrix}
     =\begin{pmatrix}
         H_J\x+\alpha\vecone_{|J|}\\
         1
     \end{pmatrix}
    \end{align*}
    As a result, we have $\hat{H}(\x+\z)\ge\alpha\vecone$ (followed from $H_J\x\ge\veczero$), $\vecone^{\transpose}(\x+\z)=1$, and $(\x + \z) \ge \gamma \vecone$. 

 \paragraph{$\z$ satisfies constraints (1) outside of $J$}
Now we will show that the $\z$ found above satisfies $\hat{H}_{\setminus J} (\x + \z) \geq \alpha$ using the fact that $\x$ satisfied $H_{\setminus J} \x \geq \rtwo\cdot\vecone$.
\begin{align*}
    \hat{H}_{\setminus J}(\x+\z) &= (H_{\setminus J} + E_{\setminus J})(\x+\z) \tag{where $E_{\setminus J} = \hat{H}_{\setminus J} - {G}_{\setminus J}$} \\
    &\geq \rtwo\cdot\vecone + E_{\setminus J}(\x+\z) + H_{\setminus J}\z 
    \tag{$H_{\setminus J}\z\x\ge\rtwo\vecone$}
    \\
    &\geq \left(\rtwo - \|E_{\setminus J}(\x+\z)\|_\infty - \|H_{\setminus J}\z\|_\infty\right)\cdot\vecone \\
    &\geq \left(\rtwo - \|E_{\setminus J}\|_\infty \|\x+\z\|_1 - \sqrt{m+n} \|H_{\setminus J}\|_1 \|\z\|_2\right)\cdot\vecone \\
    &\geq \left(\rtwo -\alpha - \sqrt{mn(m+n)} \cdot \|\z\|_2\right)\cdot\vecone
\end{align*}

The constraints in $\hat{H}_{\setminus J}$ are satisfied
when $\|\z\|_2\le \frac{\rtwo-\alpha}{\sqrt{mn(m+n)}}$, which we will show next.

\paragraph{Upper bounding $\|\z\|_2$.}
Since $\z = \hat{M}^{\dagger}
         \begin{pmatrix}
         E_J \x + \alpha\vecone_{|J|} \\
         0
     \end{pmatrix}$, we have
\begin{align*}
    \|\z\|_2 &\leq \|\hat{M}^{\dagger}\|_2\cdot
         \left\|\begin{pmatrix}
         E_J \x + \alpha\vecone_{|J|} \\
         0
     \end{pmatrix}\right \|_2
     \leq \frac{2 \alpha \sqrt{m}}{\sigma_{\min}(\hat{M})} \leq \frac{2 \alpha \sqrt{m}}{\sigmalb - \alpha \sqrt{m}} \leq \frac{4 \alpha \sqrt{m}}{\sigmalb} ,
\end{align*}
where the last step is due to the choice of sufficiently small $\alpha$, and the second-last step is due to Weyl's inequality ($\sigma_{\min}(\hat{M})\ge \sigma_{\min}(M)-\|M-\hat{M}\|_2$) and that $\sigma_{\min}(M)\ge\sigmalb$, as $M$ is a submatrix of $\augconstraints_b$ with at most $m$ rows (\Cref{assump:singular-value}). This shows that we can choose $\alpha$ to be small so that 
\[
\frac{2 \alpha \sqrt{m}}{\sigmalb - \alpha \sqrt{m}}\le \frac{\rtwo-\alpha}{\sqrt{mn(m+n)}}
\quad \Rightarrow \quad
\hat{H}_{\setminus J}(\x+\z)\ge0.
\]
The above condition holds when $\alpha=o(\frac{\sigmalb}{m^2n})$.
To conclude this part, we translate our $\ell_2$ norm bound on $\z$ into an $\ell_1$ norm bound of $\|z\|_1 \leq \sqrt{m}\|z\|_2 \leq 4 \alpha m/ \sigmalb$.
\end{proof}

Combining both lemmas, we obtain an upper bound on the $\ell_1$ Haussdorf distance between $\polytope$ and $\hat{\polytope}$ which due to the $1$-Lipschitzness of utility in $\ell_1$ norm bounds the difference in optimal utility in $\polytope$ and $\hat{\polytope}$. The proof is thus complete.

\subsection{Proof of~\Cref{lem:BinarySearchUpperBound}}
\label{app:binarysearch}
\begin{algorithm}[t]
  \caption{$\binarysearch$}
  \label{alg:binary-search}
  \Input{Left point $\x_L=\avgx^{(t-1)} \in \polytope_b$; 
         Right point $\x_R=\avgx^{(t)} \in \polytope_{b'}$; 
         step-size parameters $\varepsilon, \gamma, \alpha$.}
  \Output{Two consecutive points $\bigl(\avgx^{(t+s-1)},\,\avgx^{(t+s)}\bigr)$ on opposite sides of the boundary,
          with $\ell_1$-distance at most $\alpha$.}

  \tcp{Perform a binary search along the line segment between $\x_L$ and $\x_R$}
  $M \gets \left\lceil \log_2\!\bigl(\varepsilon/(\alpha\, t)\bigr) \right\rceil$\;

  \For{$i \gets 1$ \KwTo $M$}{
    $\z^{(i)} \gets$ target search point chosen by the binary-search rule\;
    \lIf{moving left}{ $\beta \gets \gamma$ }\lElse{ $\beta \gets \varepsilon$ }
    Keep performing $\moveonestep$ with step size $\bigl(\beta+\frac{\varepsilon}{t\,2^i}\bigr)$ toward $\z^{(i)}$\;
  }
\end{algorithm}

\begin{restatable}{lemma}{lem:BinarySearchUpperBound}
Let $\gamma$ be the minimum distance from all search point to the boundary in \Cref{lemma:estimated-polytope-close}.
Suppose the principal's average strategy moves from  $\avgx^{(t-1)}$ in polytope $\polytope_b$ to $\avgx^{(t)}$ in a different polytope $\polytope_{b'}$, with $\ell_1$ step size of $\|\avgx^{(t)}-\avgx^{(t-1)}\|_1=\frac{\eps}{t}$, where we assume $t\ge10\frac{\eps}{\gamma}$.
Then the procedure $\binarysearch$ (see \Cref{alg:binary-search}) takes at most
$s \leq O\Bigl(\frac{\varepsilon}{\gamma}+\log \frac {\varepsilon} {\alpha t}\Bigr)$
steps and returns an average strategy $\avgx^{(t+s)}$ that is within $\alpha$ $\ell_1$ distance close to the boundary
\end{restatable}

\begin{proof}
Let us refer to $\avgx^{(t)}$ as $\x_{L}$ and $\avgx^{(t-1)}$ as $\x_{R}$. Let us also refer to the direction from $\x_L$ to $\x_R$ as \emph{right} and the direction from $\x_R$ to $\x_L$ as \emph{left}. Throughout the binary search process, we move along the line between $\x_L$ and $\x_R$. For notational simplicity denote $l=\|\x_L-\x_R\|_1=\eps/t$.

Let $M$ be the number of iterations of binary search. Since each iteration halves the search space, we have $M\le \log\left(\frac{\|\avgx^{(t)}-\avgx^{(t-1)}\|_1}{\alpha}\right)=O(\log(\frac{\eps}{\alpha t}))$.

For $i\in[M]$, we denote the number of rounds spent in the $i$-th iteration as $[t_i,t_i+s_i]$, where we have $t_1=t$ and $t_{i+1}=t_i+s_i+1$. 
The total number of rounds is therefore
$s=\sum_{i=1}^M (s_i+1)=M+\sum_i s_i.$
From the halving property of binary search, we know that in iteration $i$, the total distance moved is at most $\frac{\|\avgx^{(t)}-\avgx^{(t-1)}\|_1}{2^i}$, i.e.,
\[\|\avgx^{(t_i)}-\avgx^{(t_i+s_i)}\|_1=\frac{l}{2^i}.\]
On the other hand, at each round $\tau\in[t_i,t_i+s_i)$, the average strategy $\avgx^{(t)}$ has at least ${l}/{2^i} +\min\{\eps,\gamma\}$ distance to the direction it is moving to (left or right).
From \Cref{remark:max-step-size} and the $\binarysearch$ algorithm, the maximum $\ell_1$ distance that the average strategy can travel from  $\avgx^{(\tau-1)}$ to $\avgx^{(\tau)}$ satisfies
\[\|\avgx^{(\tau)}-\avgx^{(\tau-1)}\|_1\le \frac{\frac{l}{2^i} +\min\{\eps,\gamma\}}{\tau}.\] In fact, in each step $\tau<t_i+s_i$, the principal travels exactly this distance in order to minimize the total number of steps.

Therefore, the total movement in iteration $i$ satisfies
\begin{align*}
    \frac{l}{2^i}= \|\avgx^{(t_i)}-\avgx^{(t_i+s_i)}\|_1\ge\sum_{\tau=t_i}^{t_i+s_i-1}\frac{\frac{l}{2^i} +\min\{\eps,\gamma\}}{\tau}
    \ge \left(\frac{l}{2^i} +\min\{\eps,\gamma\}\right)
    \cdot\log\left(\frac{t_i+s_i}{t_i}\right)
    \tag{$t_i\le t+s$}
\end{align*}
We can use the above inequality to upper bound $s_i$ as follows:
\begin{align*}
    s_i\le&\ t_i\left(\exp\left(\frac{\frac{l}{2^i}}{\frac{l}{2^i} +\min\{\eps,\gamma\}}\right)-1\right)\\
    \le&\ t_i\cdot\frac{l}{2^{i-1}\cdot\min\{\eps,\gamma\}}
    \tag{$e^x-1\le 2x$ when $x\in[0,1]$}\\
    \le&(t+s)\cdot\frac{l}{2^{i-1}\cdot\min\{\eps,\gamma\}}.
\end{align*}
Recall that the total number of steps is $s=\sum_i s_i+M$. Summing the above inequality over $i\in[M]$ gives us
\begin{align*}
    &s=\sum_i s_i+M \le 
    2(t+s)\cdot\frac{l}{\min\{\eps,\gamma\}}+M\\
    \Rightarrow\quad& s\lesssim \frac{\eps}{\gamma}+\log\left(\frac{\eps}{\alpha t}\right).
    \tag{From the assumption $t\ge10\frac{\eps}{\gamma}$}
\end{align*}
The proof is thus complete.
\end{proof}

\subsection{Proof of \Cref{thm:search_correctness}}\label{proof:search_correctness}
\thmsearchcorrectness*

In this section, we present the full proof of \Cref{thm:search_correctness} by expanding the proof sketch in \Cref{sec:search-polytopes-correctness}.

Let $J$ of size $j$ index the polytopes already discovered in previous iterations. We will analyze the iteration of the algorithm with the search space $\hatsubspace_J = \{\x \in \sspacep: \hat{H}_J \x \geq \alpha_j \}$, where $\hat{H}_J$ is the matrix of approximations to the hyperplanes discovered thus far. Inductively we will show that if $\|\hat{h}_{b,b_j} - {h}_{b,b_j}\| \leq \alpha_j$ for every $j \in J$, the next constructed hyperplane satisfies $\|\hat{h}_{b,\bnew} - h_{b, \bnew}\| \leq \alpha_j O(\alpha_j^2 m^2 / \sigmalb^2)$. 
We will later show how to set each $\alpha_j$ so they all remain less than $\alpha$.

\lemmaterminate*

\begin{proof}
    We will first argue that $|J| \leq m-1$. By \Cref{lemma:polytopes-far}, the set $\{\x \in \sspacep: \|H_J\x\|_\infty \leq R_2\}$ is empty if $|J| \geq m$. For any point in $\{\x \in \sspacep: \hat{H}_Jx = 0\}$, $\|H_J \x\|_\infty \leq \sqrt{m}\|\hat{H}_Jx\|_2 + \sqrt{m} \|(H_J - \hat{H}_J)\|_2 \|x\|_2 \leq m\sqrt{m} \alpha$. As long as $m\sqrt{m} \alpha$, this implies that the set $\{\x \in \sspacep: \hat{H}_Jx = 0\}$ is empty for $|J| \geq m$.
\end{proof}

\begin{lemma}[Projection of search point is close to search point]\label[lemma]{lem:search_point_close} The search point $\hat{\x}_J$ of the iteration is close to the target search point $\x^*$, where $\hat{\x}_J$ is the projection of $\x^*$ on to the search space $\hatsubspace_J$.
    \[\|\hat{\x} - \x^*\| \in O \left (\frac {\alpha_j m^2} {\sigma_{\min}(H_J) - \alpha_j \sqrt{m}} \right ). \]
\end{lemma}
\begin{proof}
  We have  
    \begin{align*}
        \|\hat{\x}_J - \x^*\| &= \|\left(I - \hat{H}_J^t (\hat{H}_J\hat{H}_J^t)^{-1} \hat{H}_J \right )\x^* - \x^*\| \\
        &= \|\hat{H}_J^t (\hat{H}_J\hat{H}_J^t)^{-1} \hat{H}_J \x^*\| \\
        &= \|\hat{H}_J^t (\hat{H}_J\hat{H}_J^t)^{-1} H_J \x^* + \hat{H}_J^t (\hat{H}_J\hat{H}_J^t)^{-1} (\hat{H}_J-H_J) \x^*\| \\
        &= \|\hat{H}_J^t (\hat{H}_J\hat{H}_J^t)^{-1} \| \left (\|H_J \x^*\| +  \|(\hat{H}_J-H_J) \x^*\| \right )\\
        &\leq m \frac{1}{\sigma_{\min}(H_J) - \alpha_j \sqrt{m}} \left (\alpha_j \sqrt{m} + \alpha m \right ) \\
        &\in O \left (\frac {\alpha_j m^2} {\sigmalb} \right ).
    \end{align*}
\end{proof}

\begin{lemma}\label{lem:appx_indifference} For all points on $\hatsubspace_J$ the agent is approximately indifferent among agent actions in $J$. 
That is for every $i,j \in J$, for all $\x \in \hatsubspace_J$, $|\ua(\x, b_i) - \ua(\x, b_j)| \leq 4 \alpha_j m$. 

\end{lemma}
\begin{proof}
    Note that for all points on $\hatsubspace_J = \{\x \in \sspacep: {H}_J \cdot \x = 0\}$, the principal is indifferent among follower actions in $J$ since each row $j \in [J]$ equality states that the utility due to agent action $b_j$ is the same as due to $b$.

    Recall that for any $b \in \cB$, $\up(\x, b) = \ip{\u_b}{\x}$.
    For any $\hat{\x} \in \{x : \hat{H}_J \cdot x = \alpha_j\}$, for any $i,j \in J$, 
    \begin{align*}
        |\ip{\u_{b_i}}{\hat{\x}} - \ip{\u_{b_j}}{\hat{\x}}| &\leq |h_i{\hat{x}} - h_j{\hat{\x}}| \\
        &\leq 2 \|{H_J}\hat{\x}\|_\infty \\
        &\leq 2 \|{H_J}\hat{\x}\|_2 \\
        &\leq 2 \|\hat{H_J} \hat{\x} + ({H_J} - \hat{H}_J)\hat{\x}\|_2 \\
        &\leq 2\alpha_j(\sqrt{m} + m).
    \end{align*}
\end{proof}

Now we will discuss how random search on the search space via Gaussian random vectors will discover new polytopes in $\mathcal{P}_{\delta}(\x^*)$ that are not yet discovered.

There are two components to this argument. The first is while there is an undiscovered polytope in $\mathcal{P}_{\delta}(\x^*)$, we will discover a new polytope. The second is that there are not many other polytopes that are discoverable. Together these two components ensure we discover all polytopes in $\mathcal{P}_{\delta}(\x^*)$.

The first component is shown in the following lemma.

\begin{lemma}\label{lemma:discover_new_poly}
Let $z$ be the random search vector in the space $\hatsubspace_J$ with step size $\eta$. That is, $z\sim \normal(\mathbf{0}^m, \eta_j^2 \phi_J^t \phi_J)$. If $J \not \supseteq \mathcal{P}_{\delta}(\x^*)$ and  $\eta_j \in \Theta(\alpha_j m^4 / \sigmalb^2)$,  then with $\Omega(1)$ probability, $\hat{x}_J + z$ does not lie in $\polytope_b$ or in any polytope $\polytope_{b'}$ for $b' \in [J]$.
\end{lemma}
\begin{proof}
    The main idea of this proof is that if there is an undiscovered polytope $b' \in \mathcal{P}_{\alpha}(\x^*)$, then \[
     \Pr_{z}\left[{h_{b,b'}}{(\hat{x}_J+z)} > \frac{\sigma_{\min}(H_J)}{\sqrt{m-1}}\cdot \eta - O\left ( \frac{\alpha m^3}{\sigma_{\min}(H_J)} \right ) - O(\delta \sqrt{m}) \right]\ge\Omega(1).
     \]  

     This is a lower bound on how much that the principal prefers action $b'$ over $b$ for $\hat{x}_J + z$. Since the principal is approximately indifferent between $b$ and any $b_j$ for $j \in J$, this is also an approximate lower bound on the principal's preference over the new action $b'$ over any already discovered action. This means that $\hat{x}_J + z$ lies in a polytope that has not already be discovered.

    We will lower bound ${h_{b,b'}}{(\x^*+z)}$ and the resulting bound follows from applying the lemma showing closeness of $\hat{\x}_J$ and $\x^*$.

    Let $j = |J|$. Let $\phi=[v_1,\ldots,v_m]^\transpose$ be a set of orthonormal vectors such that $v_{1:j}$ is the orthonormal basis of $\spanspace(H_J)$ and $v_{j+1:m}$ is the orthonormal basis of $\nullspace(H_J)$. Let $\z$ be a Gaussian vector in the null space of $H_J$, i.e., $z\sim \normal(\mathbf{0}^m,\phi_{j+1:d}^\transpose \phi_{j+1:d})$. 
    
     For any $i \notin J$, we can decompose $h_i$ into $h_i=h_i^{\parallel}+h_i^{\perp}$, where $h_i^{\parallel}\in\spanspace(H_J)$ and $h_i^{\perp}\in\nullspace(H_J)$. This decomposition gives
    \[
    \ip{h_i}{x+z}=\ip{h_i}{z}=\ip{h_i^{\perp}}{z}=\|h_i^{\perp}\|_2\cdot\ip{{h_i^{\perp}}/{\|h_i^{\perp}\|_2}}{z}.
    \]
    We will prove this lemma by showing that (1) $\|h_i^{\perp}\|_2\ge\sigma_{\min}(H_J)$; and (2) with constant probability, we have $\cos(h_i^{\perp},z)=\ip{{h_i^{\perp}}/{\|h_i^{\perp}\|_2}}{z/\|z\|_2}<-\frac{1}{\sqrt{d-j}}$.

    For the first claim, we further decompose $h_i^{\perp}=y_1+y_2$, where $y_1\in\nullspace(H_J)\cap\spanspace(H_{[n]\setminus J\setminus \{i\}})$, and $y_2\in\nullspace(H_J)\cap \nullspace(H_{[d]\setminus J\setminus \{i\}})=\nullspace(H_{[n]\setminus \{i\}})$. We have
    \begin{align*}
        \sigma_{\min}(H_J)&=\min_{y\in\br^m:\|y\|_2=1}\|H_J y\|_2\\
        &\le\frac{1}{\|y_2\|_2}\cdot\|H_J y_2\|_2 \tag{choose $y={y_2}/{\|y_2\|_2}$}\\
        &=\frac{1}{\|y_2\|_2}\cdot \left|\ip{h_i}{y_2}\right|
        \tag{$y_2\in\spanspace(H_{[d]\setminus \{i\}})$}\\
        &=\frac{1}{\|y_2\|_2}\cdot \left|\ip{y_2}{y_2}\right| \tag{$y_1\perp y_2$}\\
        &=\|y_2\|_2\le\|h_i^{\perp}\|_2.
        \tag{$\|h_i^{\perp}\|_2=\sqrt{\|y_1\|_2^2+\|y_2\|_2^2}$}
    \end{align*}

    For the second claim, we can expand $h_i^{\perp}/\|h_i^{\perp}\|$ into another set of orthonormal basis $v_{1:d-j}'$ of $\nullspace(H_J)$ and write $z=\sum_{l=1}^{d-j} v_{l}' z_l'$ where $z_{1:d-j}'\simiid \normal(0,1)$.
    Therefore, we have 
    \begin{align*}
        \cos(h_i^{\perp},z)=\ip{{h_i^{\perp}}/{\|h_i^{\perp}\|_2}}{z/\|z\|_2}=\frac{z_{1}'}{\sqrt{(z_{1}')^2+\|z_{2:d-j}'\|_2^2}}.
    \end{align*}
    Let $\delta=\frac{1}{\sqrt{d-j}}$, we have
    \begin{align*}
        \cos(h_i^{\perp},z)<-\delta&\iff z_1'<-\frac{\delta}{\sqrt{1-\delta^2}}\|z_{2:d-j}'\|_2 \\
        &\Longleftarrow z_1'<-1,\text{ and }\|z_{2:d-j}'\|_2<\frac{\sqrt{1-\delta^2}}{\delta}.
    \end{align*}
    Since $z'_{1:d-j}$ are iid Gaussian, the probability of both events are lower bounded by constants, so we have
    \begin{align*}
        \Pr[\cos(h_i^{\perp},z)<-\delta]\ge \Pr[z_1'<-1]\cdot\Pr\left[\|z_{2:d-j}'\|_2<\frac{\sqrt{1-\delta^2}}{\delta}\right]\ge\Omega(1).
    \end{align*}
    Combining the two claims proves the lemma.
\end{proof}

\begin{claim}
    With probability $\Omega(1)$ a search direction point $\z$ generated from the distribution  $\normal(\boldsymbol{0^m}, \eta_j^2 \phi_J^\transpose \phi_J)$ has $\|z\|_2 \leq O(\eta \sqrt{m})$.
\end{claim}

From the lemma and claim above, we get can show that a fraction $\Omega(1/m)$ of the search points generated lie in a new, undiscovered polytope $\polytope_{b_{\mathrm{new}}}$ and lie within distance $\eta_j\sqrt{m}$ close $\hat{\x_J}$ which in turn is $O(\alpha_j m^3 / \sigmalb)$ close to $h_{b, b_{\mathrm{new}}}$ ( and in fact any boundary in $\polytope_{\alpha}(\x^*)$). So the points in $Y$ lie at a distance $O(\eta_j \sqrt{m} + \alpha_j m^3/\sigmalb)$. From the choice of $\eta_j \in \Theta(\alpha_j m^4 / \sigma^2)$, this is $O(\alpha_j m^{4.5} / \sigmalb^2)$.

Using these properties, we will now show that with a large enough number of search points, we can approximate $h_{b, b_{\mathrm{new}}}$ well. We construct the approximation in the following way. Let $Y$ be a matrix of the search points obtained by $\hat{\x}_H + \z_J$, where $\z_J$ is a Gaussian random vector in the search space $\hatsubspace_J$. We will construct a hyperplane solving $\hat{h} = \argmin_h \|hY^\transpose\|$. We will now show that $\hat{h}$ is a good approximation to $h_{b, b_{\mathrm{new}}}$.

\begin{lemma}[Approximating a hyperplane]\label{lem:separating_hyperplane}
Suppose $\mathsf{RandomSearch}$ in the  search space $\hatsubspace_J = \{\x : \hat{H}_J\x = \alpha_j\}$ generates $d \in \Theta(m^2 \log m)$ points.
Let $\bnew$ be the undiscovered polytope with the maximum number of search points. Let $Y \in \mathbb{R}^{k \times m}$ be a matrix where the rows are all search points in $\polytope_{\bnew}$ with distance from $\hat{\x}_J \in O(\eta \sqrt{m})$.

      Then $\hat{h} = \argmin_{h} \|hY^t\|_2$ satisfies $\|\hat{h} - h_{b, b_{\mathrm{new}}}\| \leq \alpha$.  
\end{lemma}

We will prove the accuracy of $\hat{h}$ in approximating $h_{b, b_{\mathrm{new}}}$ by finding and using a lower bound on the minimum singular value of the matrix $Y$.

Let us denote the matrix of all search directions generated by random search by $Z \in \mathbb{R}^{d \times m}$. That is each row $z_i \sim \normal(\boldsymbol{0}, \eta_j^2 \phi_J^t \phi_J)$. Let $W \in \mathbb{R}^{k \times m}$ denote the subset of rows of $Z$ corresponding to search vectors with length at most $O(\eta_j \sqrt{m})$ and lying in $\polytope_{\bnew}$. $W$ has a $\Omega(1/m)$ fraction of the rows in $Z$.  The matrix of close-to-boundary search points in $\polytope_{\bnew}$ is $Y$ where each row $y_i$ is $\hat{\x}_J + w_i$.

The minimum singular value of $Z \geq \eta$.  
Since $W$ has $\Omega(1/m)$ fraction of subsets of $Z$, we can show that the minimum singular value of $W$ is also lower bounded. We show that $ \sigma_{\min}(W) \in \Omega(\eta_j \sqrt{m / d})$ which is $\Omega(\eta_j \sqrt{1/(m \log m)})$ by our choice of $d$.

Each row of $Z$ is an isotropic random vector in a dimension $p \leq m$. Fix some $u \in \mathbb{R}^p$. Then, each entry of $Zu$ is normally distributed $N(0,1)$, i.i.d. With high probability, at least an $1/m$-fraction of the entries will be at least $1/(2m)$ in absolute value, from concentration. Now, we take a union bound over an $\epsilon$-Net over the set of all unit vectors. With high probability, this will hold for any vector in the net. If this holds for any vector of the net, we can say that this holds for all unit vectors and the proof would be complete.

Next note that $\sigma_{d-1}(Y) \geq \sigma_{d}(W)$ since $Y-W$ is a matrix of rank 1. 

Armed with the property that $\sigma_{\min}(Y) \geq \Omega(\eta_j \sqrt{1/m \log m})$, we will now show that $\hat{h} = \argmax_{h} \|hY^t\|_2$ is a good approximation for $h_{b, b_{\mathrm{new}}}$.

The true boundary $h_{b, b_{\mathrm{new}}}$ also has a small $\|h_{b, b_{\mathrm{new}}} Y^t\|_2 \leq O(\alpha_j m^{5.5} / \sigmalb^2)$ since points in $Y$ are close to the boundary. Let $v$ denote the singular vector of $Y^t$ corresponding to the smallest singular value, where $\|v\|_2=1$. Write $\hat{h} = \cos(\theta)v+\sin(\theta)u$ for $u$ perpendicular to $v$. Then,
\[
\|\hat{h}Y^t\|^2 = \cos^2(\theta) \|vY^t\|^2 + \sin^2(\theta) \|uY^t\|^2 \ge \sin^2(\theta) \sigma_{d-1}(Y).
\]
Since $\|\hat{h}Y^t\|^2 \leq O(\alpha_j^2 m^{9} / \sigmalb^4)$ is small, this means that $\sin(\theta)$ is small. $\sin^2(\theta) \leq O(\alpha_j m^{6} / \sigmalb^2)$. Hence $\|\hat{h}-v\| \precsim 2 \sin \theta \in O(\sqrt{\alpha_j} m^3 / \sigmalb)$ is small. Similarly, $\|h_{b, b_{\mathrm{new}}}-v\|$ is small. Hence $\|h_{b, b_{\mathrm{new}}}-\hat{h}\| \in O(\sqrt{\alpha_j} m^3 / \sigmalb)$ is small.

We have inductively shown how to find approximations of accuracy $\alpha_j$ for the $j^{\text{th}}$ discovered hyperplane where $\alpha_{j-1} \in O(\alpha m^3 / \sigmalb) \alpha_j$. To ensure that all hyperplanes are approximated to within $\alpha$ level, we set $\alpha_1$ so that $\alpha_1 (m^3 / \sigmalb)^m \leq \alpha$ or $\alpha_1 \in O(\alpha \cdot (\sigmalb / m^3))^m$. 

\paragraph{Guarantee that all polytopes in $\mathcal{P}_{\delta}$ are discovered.} \Cref{lemma:discover_new_poly} shows that the algorithm keeps finding a new polytope as long as the search space is non-empty and some polytope in $\mathcal{P}_{\delta}(\x^*)$ is not yet visited. The only way the algorithm can terminate without finding some polytope in $\mathcal{P}_{\delta}(\x^*)$ is if it finds $m$ polytopes that are not a superset of $\mathcal{P}_{\delta}(\x^*)$.

All the polytopes discovered have points at distance $\alpha_j m^3 / \sigmalb$ from $\x^*$ and hence are polytopes in $\mathcal{P}_{\alpha_j m^3 / \sigmalb}(\x^*)$. By choosing $\alpha_j$'s so that each ${\alpha_j m^3 / \sigmalb} \in \omega(R_2)$, we have the property that all the discovered polytopes belong to the set $\mathcal{P}_{R_2}(x^*)$. By \Cref{lemma:polytopes-far}, there are at most $m$ polytopes in $\mathcal{P}_{R_2}(x^*)$.

\subsection{Proof of \Cref{thm:lower_bound}}
\label{proof:lower_bound}

\thmlowerbound*

\begin{proof}
[Proof of \Cref{thm:lower_bound}]
Let $\ell$ be the largest integer such that $1/(2\ell+1) \ge \epsilon$. We can assume that the optimizer has $n=\ell^2$ actions and the learner has $m=\ell^3+1$ actions: if they have more actions then it is easy to guarantee that they are not played, by constructing a suitable game. We will prove the version without the smoothed-analysis. However, it is clear from the proof that it still holds even when the utilities are perturbed as claimed above.

Assume that the optimizer has $n$ actions. We want to define the optimizer's utility such that, in order to find a local optima, a lot of time has to be passed until the optimizer succeeds. We notice that the optimizer's utility is defined on the simplex $\Delta_n$, and it is piecewise linear: the simplex is split into polytopes, defined by the learner's best responses. For intuition, we explain now how these polytopes look like. First, there's some path of length $\ell$, where $\ell < n$, consisting of distinct vertices $v_1-\cdots v_\ell$, where each $v_i \in [n]$, and $v_1=1$. The optimizer doesn't know the path, and the only local optima of the function would be if the optimizer plays deterministically the action $v_\ell$. The optimizer would have to find $v_\ell$ in order to find a local optima, and this will take a long time. Here is how we define the polytopes such that the only local optima is $v_\ell$. First, for each edge in the path, there's some polytope: namely, for each $i=1,\dots,\ell-1$, there's a polytope around the edge connecting $v_i$ with $v_{i+1}$. We denote this polytope $P_i$ and it is approximately defined as 
\[
P_i = \{x \in \Delta_n \colon x_{v_i}+x_{v_{i+1}} \ge \max\{0.9,x_{v_{i-1}}+x_{v_i},x_{v_{i+1}}+x_{v_{i+2}}\}~\}
\]
In the remainder of the the simplex, there will be the polytope $P_0$ defined as
\[
P_0 = \{x \in \Delta(n) \colon \forall i\in[\ell-1],~x_{v_i}+x_{v_{i+1}}\le 0.9 \}
\]
The optimizer's utility is defined such that for $i<j \in \{0,\dots,\ell\}$ it is higher in $P_j$ compared to $P_i$. Within the polytopes the optimizer's utility is as follows: in $P_0$ the utility is higher as $x$ approaches $v_1=1$ (i.e. approaches the pure action $v_1=1$). For $i>1$, the utility is higher as $x$ approaches $v_{i+1}$. Concretely, the learner's utility is defined as follows: In $x \in P_i$ it equals $(2i+x_{v_{i+1}})/(2\ell+1)$.
It is trivial to see that there's not $1/(2\ell+1)$-approximate local Stackalberg except for $x$ in the vicinity of $v_\ell$.
Indeed, within each $P_i$ the local opt is only $v_{i+1}$. And further, for each $i<\ell$, $v_i$ is not a local opt because $v_i$ borders the polytope $P_i$ whose local optima is $v_{i+1}$. Consequently, the only local opt is $v_\ell$. 

Now, we define the optimizer's actions and the utility functions such that the optimizer's utility is as defined above. Recall that the optimizer knows their own utility, however, they don't know the path. Consequently, for any potential edge $v_i-v_{i+1}$ there should be a potential polytope. Some of the potential polytopes will not exist, however, the optimizer would not know that in advance. Concretely, for each ``potential polytope'' there would be an action of the learner. First, for the big polytope $P_0$, there will be an action of the learner that we denote as $0$. Further, for any $r \in [n]$, $s \in [n]\setminus \{r\}$ and $i=1,\dots,\ell$, we define an action of the learner $(r,s,i)$. We want to ensure the following:
\begin{itemize}
    \item If $v_i=r$ and $v_{i+1}=s$ then the polytope $P_i$ will correspond to the optimizer playing action $(r,s,i)$. Otherwise, this action will correspond to no polytope.
    \item If this action does correspond to $P_i$ then we want the optimizer's utility to be as defined above.
\end{itemize}
To achieve the following, we first define the learner's utility. For action $0$ of the learner, and any action $a$ of the optimizer, define
\[
u_2(a,0) = 0~.
\]
This corresond's to the ``default'' polytope $P_0$. For any $(r,s,i)$ such that $v_i\ne r$ or $v_{i+1}\ne s$, we want to ensure that there's no polytope corresponding to this action, hence we define
\[
u_2(a,(r,s,i)) = -1
\]
for any action $a$ of the optimizer, and this ensures that the learner will always prefer action $0$ over $(r,s,i)$. Next, we want to define the learner's utility on $(r,s,i)$ such that $v_i=r$ and $v_{i+1}=s$ such that the region when the learner plays $(r,s,i)$ corresponds to the polytope $P_i$. We define the learner's utility as 
\[
u_2(a,(r,s,i)) = \begin{cases}
    1/9 & a \in \{r,s\}\\
    -1 & a \notin \{r,s\}
\end{cases}
\]
This definition of $v_L$ yields the polytopes $P_i$ as defined above, namely, $P_0$ is the region where the learner plays action $0$ and for $i>0$, $P_i$ is exactly the region where learner plays $(v_i,v_{i+1},i)$.

Now, we need to define the optimizer's utility to align with the definitions above. Recall that we want the optimizer's utility to be $(x_{v_{i+1}}+2i)/(2\ell+1)$ in each region $P_i$. Starting at $P_0$, we define:
\[
u_1(a,0) = \begin{cases}
  1/(2\ell+1) & a=v_1\\
  0 & a\ne v_1
\end{cases}
\]
For $i>0$, define
\[
u_1(a,(v_i,v_{i+1},i)) = \begin{cases}
    (1+2i)/(2\ell+1) & a = v_{i+1}\\
    2i/(2\ell+1) & a \ne v_{i+1}
\end{cases}
\]

For all $(r,s,i)$,
\[
u_1(a,(r,s,i)) = \begin{cases}
    (1+2i)/(2\ell+1) & a = s\\
    2i/(2\ell+1) & a \ne s
\end{cases}
\]
It is easy to see that this yields the correct utility function. 

Now, we want to prove that it takes a long time for the optimizer to get to $v_\ell$. First, recall that $\avgx^{(t)}$ is the average optimizer's history till time $t$. Notice that $\|\avgx^{(t)} - \avgx^{(t+1)}\|_1 \le 1/(t+1)$. Denote by $e_1,\dots,e_k$ the indices of the polytopes visited throught the algorithm, except for polytope $P_0$: $e_1$ is the first visited polytope, $e_2$ is the second etc. Denote by $t_i$ the first time that $e_i$ was visited such that $t_1 \ge 1$. We want to argue that $t_{i+3} \ge 1.8 t_i$. Indeed, notice that each $P_i$ has two neighboring polytopes (except for $P_0$): $P_{i-1}$ and $P_{i+1}$. Consequently, one of $P_{e_{i+1}},P_{e_{i+2}},P_{e_{i+3}}$ does not neighbor $P_{e_i}$. Denote by $0=t_0<t_1<\cdots<t_k$ the iterations such that for $i\ge 1$: $t_i$ is the first $t>t_{i-1}$ such that $\avgx^{(t)}$ has some coordinate that's is greater than $0.45$ for the first time, namely, such that there's some $j$ such that $(\avgx^{(t)})_j \ge 0.45$ and $\avgx^{(t')}_{j}<0.45$ for all $t'<t$. We want to prove that $t_{i+2} \ge 1.35 t_i$. To show that, first, by definition, either the total variation between $\avgx^{(t_i)}$ and $\avgx^{(t_{i+1}}$ is at least $0.175$ or between $\avgx^{(t_i)}$ and $\avgx^{(t_{i+2}}$: indeed, in $\avgx^{(t_i)}$, $\avgx^{(t_{i+1}}$ and $\avgx^{(t_{i+2}}$ there are distinct coordinates that surpass $0.45$ for the first time: for one of $i+1$ and $i+2$ this coordinate has to have a weight of at most $0.55/2=0.275$ at time $t=t_i$, which implies a total variation of at least $0.45-0.275=0.175$. Further, notice that the total variation between $\avgx^{(t)}$ and $\avgx^{(t+1)}$ is at most $1/(t+1)$ by definition of $\avgx^{(t)}$ (it is the average of everything up to time $t$. Consequently, it takes at least $0.175t_i$ iterations until reaching some $\avgx^{(t)}$ whose total variation is $0.175$ from $t_i$, which means that $t_{i+2}\ge 1.175 t_i$ as claimed. This implies that the number of iterations is $e^{\Omega(k)}$ and we would like to lower bound $k$. 

Denote by $u_i$ the coordinate that's at least $0.45$ for the first time at $t_i$. We assume that the algorithm stops when reaching close to a local optima which is at $v_{\ell}$ hence there must be some $i \le k$ such that $u_i = v_\ell$. 
Denote by $I$ the set of indices $i$ such that $u_i$ is not on the path, namely
\[
I = \{ i \colon u_i \notin \{v_1,\dots,v_\ell\}\}
\]
Further, denote by $J$ the set of times $i$ that $u_i \ne v_1$ is on the path but no neighbor of $u_i$ was visited before:
\[
J = \{ i \colon \exists j>1 \text{ s.t. } u_i = v_j \text{ and } \forall i'<i, u_{i'} \notin \{v_{j-1},v_{j+1}\} \}
\]
Now, we consider $I$ and $J$ as random variables. The randomness is both wrt the randomness of the optimizer's algorithm and wrt to a random choise of a path $v_1-\cdots-v_\ell$ which is taken uniformly at random from all paths starting at $v_1=1$ whose all vertices are distinct and whose elements are $[n]$. Whenever $J = \emptyset$, then, we have that $k \ge \ell$: indeed, recall that $v_\ell \in \{u_1,\dots,u_k\}$ and if $J = \emptyset$ this means that all vertices $v_1,\dots,v_{\ell-1}$ have to appear w.p. 0.45 before $v_\ell$. So, if $\Pr[|J|=\emptyset \ge 0.5]$ then we have that the expected number of iterations run by our algorithm is at least $e^{\Omega(k)}\ge e^{\Omega(\ell)}$.

Now assume otherwise, that $\Pr[|J|\ne\emptyset ]\ge 0.5$. In this case, w.p. 0.5, there's some node $u_i$ such that no neighbor has appeared before w.p. 0.45. Since the path is uniformly at random, $\mathbb{E}[|I|] \ge \Omega(n/\ell) \mathbb{E}[|J|]$: whenever that algorithm visits a node that none of its neighbors appeared on the path, the probability that this new node is on the path is at most $O(n/\ell)$, assuming that $n>10\ell$. Hence, in case that $\E[|J|]\ge 0.5$ we have that $\E[|I|]\ge \Omega(n/\ell)$ and $\E[k] \ge \Omega(n/\ell)$. If we assume that $n \ge \ell^2$, this is $e^{\Omega(\ell)}$, as required.

\end{proof}

\subsection{Proof of \Cref{thm:smoothed-analysis}}
\label{app:smooth}
\thmsmoothedanalysis*

\begin{proof}[Proof of \Cref{thm:smoothed-analysis}]
    We first show that it suffices to establish a lower bound on the minimum singular values of all $k\times k$ submatrices of the \emph{un-augmented constraint matrices} $\constraints_b$, where $k\le m$. In particular, we will show that $\forall b\in\aspacea$,
    \begin{align}
        \Pr\Big(
        \forall K\in\sqsubmatrix_{m}(\augconstraints_b),
        \ \sigma_{\min}(K)\ge\sigmalb\Big)
        \ge \Pr\Big(
        \forall K'\in\sqsubmatrix_{\le m}(\constraints_b),
        \ \sigma_{\min}(K')\ge m\cdot\sigmalb\Big)
        \label{ineq:all-square-submatrices}
    \end{align}
    Let $\augconstraints_b\setminus=\begin{bmatrix}
        I_m\\ \vecone_m
    \end{bmatrix}$ be the augmented constraints to account for the simplex constraints $\x\ge\veczero$ and $\vecone^{\transpose} \x=1$.
    We establish \Cref{ineq:all-square-submatrices} by proving the following two claims:
    \begin{enumerate}
        \item If a square submatrix $K\in\sqsubmatrix_m(\augconstraints_b)$ contains $r$ rows from the nonnegativity constraints $I_m$, then there exists a $(m-r)\times(m-r)$ submatrix $K'\in \sqsubmatrix_{m-r} (K)$, such that $\sigma_{\min}(K)\ge\frac{1}{2}\sigma_{\min}(K')$;
        \item For a submatrix $K\in\sqsubmatrix_m(\augconstraints_b)$ that contains the row $\vecone_m$, its $\sigma_{\min}(K)$ can be viewed as a Gaussian perturbed matrix.
    \end{enumerate}
    \paragraph{Proof of the first claim.} Without loss of generality, we can assume the nonnegativity constraints are located in the first $r$ rows and first $r$ columns of $K$, i.e., $K$ takes the following form,
    \begin{align*}
        K=\begin{bmatrix}
        I_{r\times r} & \veczero_{r\times(m-r)}\\
        L & K'
\end{bmatrix},
    \end{align*}
    where $L\in\br^{(m-r)\times r}$ and $K'\in\br^{(m-r)\times (m-r)}$ is a square sub-matrix of $K$. We will show that $\sigma_{\min}(K)\ge\frac{1}{m}\sigma_{\min}(K')$ by proving that for all $m$-dimensional vector 
    $\x\in\br^{m}$ where $\|\x\|_2=1$, we have $\|K\x\|_2\ge \frac{1}{m}\sigma_{\min}(K')$.

    We can write $\x=\begin{bmatrix}
        \y\\ \z 
    \end{bmatrix},$ where $\y\in\br^{r}$ and $\z\in\br^{n-r}$. We have $K\x=\begin{bmatrix}
        \y\\
        L\y+K'\z
    \end{bmatrix}$. Consider the following two cases:
    \begin{itemize}
        \item If $\|\y\|_2\ge\frac{1}{m}\cdot\sigma_{\min}(K')$, then $\|K\x\|_2=\|\y\|_2+\|L\y+K'\z\|_2\ge\|\y\|_2\ge \frac{1}{m}\sigma_{\min}(K')$, as desired.
        \item If $\|\y\|_2<\frac{1}{m}\cdot\sigma_{\min}(K')$, we have $\|\z\|_2\ge1-\frac{1}{m}$. In this case,
        \begin{align*}
            \|K\x\|_2=&\ \|\y\|_2+\|L\y+K'\z\|_2
        \ge \|\y\|_2+\|K'\z\|_2-\|L\y\|_2\\
        \ge&\ \sigma_{\min}(K')\cdot\|\z\|_2-(\|L\|_2-1)\cdot\|\y\|_2\\
        \ge&\ \sigma_{\min}(K')\cdot \left(
        1-\|\y\|_2\right)
        -\left(\frac{m}{2}-1\right)\cdot\|\y\|_2
        \tag{$\|L\|_2\le \sqrt{r(m-r)}\le\frac{m}{2}$}\\
        \ge&\ \sigma_{\min}(K')\cdot\left(
            1-\frac{\sigma_{\min}(K')}{m}
            -\left(\frac{m}{2}-1\right)\cdot\frac{1}{m}
        \right)
        \tag{$\|\y\|_2\le \sigma_{\min}(K')/m$}\\
        =&\ \sigma_{\min}(K')\cdot\left(
            \frac{1}{2}+\frac{1-\sigma_{\min}(K')}{m}
        \right)\\
        \ge&\ \frac{\sigma_{\min}(K')}{2}\ge \frac{\sigma_{\min}(K')}{m}.
        \end{align*}
    \end{itemize}
    This finishes the proof of the first claim.

    \paragraph{Proof of the second claim.}
    Let $K=\begin{bmatrix}
        \vecone_m\\
        K'
    \end{bmatrix}$ be a submatrix of $\augconstraints_b$ that contains the all-1 row vector.
    Since the Gaussian distribution is rotation invariant, we can rotate all the rows of $\boldsymbol{W}$ to $V\cdot\boldsymbol{W}$, where $V$ is a rotation matrix. The resulting matrix $V\boldsymbol{W}$ satisfies that (1) the first row equals $c\cdot\vecone_m$, where $c\sqrt{m}$ is the length of the first row that follows the Chi distribution with $m$ degrees of freedom; and (2) the distribution of the remaining $m-1$ rows does not change, i.e., their entries remains to be i.i.d. Gaussian random variables.
    We can therefore view $K=\begin{bmatrix}
        \vecone_m\\
        K'
    \end{bmatrix}$ as some fixed matrix perturbed by a Gaussian matrix with variance $c^2\sigma^2$, where $c^2\ge1-O\left(\sqrt{\frac{\log(1/\delta)}{m}}\right)$ with probability at least $1-\delta$. 
    As a result, the minimum singular value of $K$ also follows from the characterization in \Cref{lemma:gaussian-perturbed}. 
    
    \paragraph{Union bound on all submatrices}
    Finally, we are ready to bound the tail probability of the minimum singular value. From \Cref{ineq:all-square-submatrices}, it suffices to consider all the square submatrices of $\augconstraints_b$ with size at most $m$, and show that they all have minimum singular value lower bounded by $\sigmalb m$. From \Cref{lemma:gaussian-perturbed}, for a given $b\in\aspacea$, $r\le m$ and $K\in\sqsubmatrix_{r}(\constraints_b)$, we have
    \begin{align*}
        \Pr\left(\sigma_{\min}(K)\le \sigmalb m\right)\le
        O\left(
        \frac{m^{3/2}}{\sigma}\cdot\sigmalb
        \right).
    \end{align*}
    Therefore, taking a union bound for all such submatrices, we have
    \begin{align*}
        \Pr\Big(
        \exists b\in\aspacea, r\le m, K'\in\sqsubmatrix_{r}(\constraints_b),
        \ \sigma_{\min}(K')\ge m\cdot\sigmalb\Big)
        \le O\left(
        \frac{m^{\frac{5}{2}}}{\sigma}\cdot\sigmalb\cdot
        \binom{n}{\le m}
        \right)
        \le O\left(
        \frac{m^{\frac{5}{2}}2^n}{\sigma}\cdot\sigmalb
        \right).
    \end{align*}
    Finally, setting the right hand side probability to be $\delta$, we have established that
    \begin{align*}
        \Pr\left(\forall b\in\aspacea,
        \forall K\in\sqsubmatrix_{m}(\augconstraints_b),
        \ \sigma_{\min}(K)\ge
        \frac{\sigma\delta}{m^{\frac{5}{2}}2^n}
        \right)
        \ge1-\delta.
    \end{align*}
    The proof is thus complete.
\end{proof}

\begin{lemma}[Theorem 3.3 of \citep{sankar2006smoothed}]
\label[lemma]{lemma:gaussian-perturbed}
    Let $\boldsymbol{\overline{A}}\in \br^{m\times m}$ be an arbitrary square matrix, and let $\boldsymbol{A}$ be a Gaussian perturbation of $\boldsymbol{\overline{A}}$ of variance $\sigma^2$. Then 
    \begin{align*}
        \Pr\left(\sigma_{\min}(\boldsymbol{A})\le x\right)\le 2.35\frac{\sqrt{m}}{\sigma}\cdot x.
    \end{align*}
\end{lemma}

\end{document}